\documentclass[fleqn,usenatbib]{mnras}

\usepackage{newtxtext,newtxmath}
\usepackage{amsmath}
\newcommand{\angstrom}{\text{\normalfont\AA}}

\usepackage[T1]{fontenc}
\usepackage{ae,aecompl}

\usepackage{graphicx}	
\usepackage{amsmath}	
\usepackage{amssymb}	

\title[Probing Cosmic Dawn with Emission Lines]{Probing Cosmic Dawn with Emission Lines: Predicting Infrared and Nebular Line Emission for ALMA and JWST}

\author[H. Katz et al.]{Harley Katz$^{1}$\thanks{harley.katz@physics.ox.ac.uk}, Thomas P. Galligan$^1$, Taysun Kimm$^2$, Joakim Rosdahl$^3$, \newauthor Martin G. Haehnelt$^4$, Jeremy Blaizot$^3$, Julien Devriendt$^1$, Adrianne Slyz$^1$, \newauthor Nicolas Laporte$^5$, \&  Richard Ellis$^5$
\\
$^1$Astrophysics, University of Oxford, Denys Wilkinson Building, Keble Road, Oxford OX1 3RH, UK \\
$^2$Department of Astronomy, Yonsei University, 50 Yonsei-ro, Seodaemun-gu, Seoul 03722, Republic of Korea \\  
$^3$Univ Lyon, Univ Lyon1, Ens de Lyon, CNRS, Centre de Recherche Astrophysique de Lyon UMR5574, F-69230, Saint-Genis-Laval, France \\
$^4$Kavli Institute for Cosmology and Institute of Astronomy, Madingley Road, Cambridge CB3 0HA, UK \\
$^5$Department of Physics \& Astronomy, University College London, London, WC1E 6BT, UK
}

\date{Accepted XXX. Received YYY; in original form ZZZ}

\pubyear{2015}

\begin{document}
\label{firstpage}
\pagerange{\pageref{firstpage}--\pageref{lastpage}}
\maketitle

\begin{abstract}
Infrared and nebular lines provide some of our best probes of the physics regulating the properties of the interstellar medium (ISM) at high-redshift.  However, interpreting the physical conditions of high-redshift galaxies directly from emission lines remains complicated due to inhomogeneities in temperature, density, metallicity, ionisation parameter, and spectral hardness.  We present a new suite of cosmological, radiation-hydrodynamics simulations, each centred on a massive Lyman-break galaxy that resolves such properties in an inhomogeneous ISM.  Many of the simulated systems exhibit transient but well defined gaseous disks that appear as velocity gradients in [CII]~157.6$\mu$m emission.  Spatial and spectral offsets between [CII]~157.6$\mu$m and [OIII]~88.33$\mu$m are common, but not ubiquitous, as each line probes a different phase of the ISM.  These systems fall on the local [CII]-SFR relation, consistent with newer observations that question previously observed [CII]~157.6$\mu$m deficits.  Our galaxies are consistent with the nebular line properties of observed $z\sim2-3$ galaxies and reproduce offsets on the BPT and mass-excitation diagrams compared to local galaxies due to higher star formation rate (SFR), excitation, and specific-SFR, as well as harder spectra from young, metal-poor binaries.  We predict that local calibrations between H$\alpha$ and [OII]~3727$\angstrom$ luminosity and galaxy SFR apply up to $z>10$, as do the local relations between certain strong line diagnostics (R23 and [OIII]~5007$\angstrom$/H$\beta$) and galaxy metallicity.  Our new simulations are well suited to interpret the observations of line emission from current (ALMA and HST) and upcoming facilities (JWST and ngVLA).
\end{abstract}

\begin{keywords}
galaxies: high-redshift, galaxies: ISM, dark ages, reionization, first stars, ISM: lines and bands, ISM: kinematics and dynamics, galaxies: star formation 
\end{keywords}



\section{Introduction}
\label{intro}
Understanding stellar mass assembly and the evolution of the properties of the interstellar medium (ISM) across cosmic time remain some of the most fundamental goals in modern astrophysics.  One of the primary methods of achieving this is through the study of emission lines at various wavelengths that directly probe different phases of the ISM as well as the different stellar populations that may be responsible for exciting them.  For example, forbidden infrared (IR) lines from [NII]~121.5$\mu$m and [OIII]~88.3$\mu$m trace ionised gas, while other lines, such as [CII]~157.6$\mu$m, trace the neutral ISM \citep{Maiolino2015}.  Combining the information from each individual IR line provides a detailed picture of the properties of the ISM.  In the low-redshift Universe, state-of-the-art observational facilities, such as the {\it Herschel Space Observatory} \citep{Pillbratt2010}, have made it possible to target many different lines in the same galaxy \citep[e.g.][]{Kennicutt2011}.  The results of these observational campaigns can be interpreted with dust radiative transfer models to elucidate important characteristics of the galaxy including dust mass, dust production mechanisms, and star formation \citep[e.g.][]{DeLooze2016}.  

In addition to IR lines, nebular emission lines from either recombining or collisionally excited gas have been the workhorse for determining ISM properties such as gas temperature, electron density, chemical composition, and ionisation parameter \citep[e.g.][]{Evans1985}.  Likewise, nebular line luminosities from H$\alpha$ and the [OII]~3727$\angstrom$ doublet have been the primary measures of the star formation rates (SFRs) in galaxies \citep{Kennicutt1998a}.  Further applications include classifying starburst galaxies from active galactic nuclei (AGN) \citep[e.g.][]{Kewley2001,Kauffmann2003,Kewley2006}.  The development of integral field spectroscopy (IFS) has opened the opportunity to go beyond categorising galaxies by their global line ratios, and can constrain how the emission line properties change between individual regions or molecular clouds within the same system \citep{Bacon1995,Bacon2001}.  A particularly interesting use of IFS has been to spatially resolve classic line diagnostic diagrams \citep[e.g.][]{Belfiore2016}, such as the Baldwin-Philips-Terlevich (BPT) diagnostic \citep{Baldwin1981}, to identify the differences between various galactic regions, such as the nucleus and disk. 

In the low redshift Universe, fine structure lines, including [CII]~157.6$\mu$m and [OIII]~88.3$\mu$m, are likely the dominant coolants of the ISM \citep{Spitzer1978} and hence are also expected to be bright in the early Universe, once galaxies have been sufficiently metal enriched.  ALMA has been instrumental in detecting these emission lines in the early Universe, even in the more ``normal" galaxies with SFRs of $\sim10$M$_{\odot}$yr$^{-1}$ \citep[e.g.][]{Maiolino2015,Carniani2017}.  [OIII]~88.3$\mu$m has even been detected at $z>9$ \citep{Hashimoto2018}.  These IR studies have revealed numerous unexpected features in the high-redshift galaxy population such as spatial and spectral offsets between different IR emission lines, the UV, and Ly$\alpha$ \citep[e.g.][]{Carniani2017}, possible ordered rotation \citep{Smit2018}, and deficits in [CII]~157.6$\mu$m luminosity compared to systems in the local Universe \citep[e.g.][]{Ouchi2013}.

Compared to IR emission lines, detections of individual nebular emission lines in the epoch of reionization has been limited because at $z\gtrsim4$, emission lines in the rest-frame optical are redshifted to the mid-infrared where they are difficult to observe from the ground.  For this reason, rest-frame optical lines at high redshift are a primary target for the {\it James Webb Space Telescope} (JWST).  Although they are currently difficult to detect, their presence can be inferred in the broad-band filters of current space-based facilities such as {\it Spitzer} \citep[e.g.][]{Stark2013}.  Our current, best categorised samples of nebular emission line galaxies exist at $z\sim2-3$ \citep[e.g.][]{Steidel2014}.  Similar to the IR studies at high redshifts, these $z\sim2-3$ galaxies also exhibit interesting features such as offsets on the BPT diagram with respect to low-redshift SDSS galaxies, super-solar ratios of O/Fe, and a high degree of nebular excitation with respect to low-redshift systems \citep{Steidel2016,Strom2017}.  

Understanding the changing behaviour of nebular, and IR emission at high redshifts is a key theoretical problem.  Historically, photoionisation codes, such as {\small CLOUDY} \citep{Ferland1998} have been a primary method for generating the nebular and IR emission line luminosities for a specific set of ISM conditions under the assumption of an impinging flux with a given spectral shape.  However, in order to accurately model a realistic galaxy, one requires a detailed understanding of the properties of the ISM, gas distribution and kinematics, and radiation field across the entire system.  Numerical simulations of galaxy formation with a fully coupled radiative transfer code are thus an ideal tool to use for this purpose.

Recently, there has been an increased effort in using high resolution cosmological simulations to model the emission properties of high-redshift galaxies to better understand and complement the observations \citep{Zackrisson2013,Cen2014,Wilkins2016,Vallini2015,P17a,P17b,Katz2017,Barrow2017,Moriwaki2018,Arata2018,Smith2018}.  It is currently computationally unfeasible to resolve all of the necessary scales or physics needed to calculate much of this emission from first principles.  However, we argue that at minimum, we need the combination of a resolved multi-phase ISM, coupled radiation-hydrodynamics to model an inhomogeneous radiation field and hydrogen ionisation state, and realistic cooling and feedback prescriptions.  Such simulations include the Renaissance suite \citep{Oshea2015} which has been used by \cite{Barrow2017} to compute the spectra of high-redshift galaxies, or the SPHINX simulations \citep{Rosdahl2018}.  However, these simulations do not yet reach the combination of galaxy mass, redshift, and number of radiation bands needed to directly comparable with observed high-redshift galaxies.

In this work, we have developed a new suite of simulations, called the Aspen Simulations, which are cosmological radiation-hydrodynamics simulations specifically designed to predict IR and nebular emission line luminosities for high-redshift galaxies.  The goal of these simulations is {\it not} to specifically match {\it all} of the observational properties of high-redshift galaxies, but rather to better understand the physics that is governing the ISM when we do match the observations and to further identify shortfalls in our modelling when our simulations do not agree with observations.  To this end, we present a new framework which allows us to both quickly and accurately predict IR and nebular emission lines from cosmological simulations.  

This paper is organised as follows: In Section~\ref{methods}, we describe the Aspen simulations and our newly developed technique for estimating IR and nebular emission lines from cosmological simulations.  In Sections~\ref{results_GP} and~\ref{results_OC}, we address the two sets of emission lines independently and directly compare the predictions from our simulations to multiple observational probes of the epoch of reionization. Finally, in Section~\ref{dc}, we present the caveats associated with our work as well as our discussion and conclusions.

\section{Methods}
\label{methods}
The simulations used in this work were first presented in \cite{Katz2018}.  We model the formation of the region around a massive, high-redshift Lyman-break galaxy (LBG), using a cosmological gravitational radiation-hydrodynamics zoom-in simulation.   We exploit the publicly available, adaptive mesh refinement (AMR) code {\small RAMSES-RT} \citep{Teyssier2002,Rosdahl2013,Rosdahl2015}.

\subsection{Initial Conditions}
We begin by running a low-resolution ($256^3$) dark-matter only simulation in a cosmological box with side length 50~comoving~Mpc.  Initial conditions were generated with {\small MUSIC} \citep{Hahn2011} using a \cite{Planck2016} cosmology ($h=0.6731$, $\Omega_{\rm m}=0.315$, $\Omega_{\rm b}=0.049$ , $\Omega_{\Lambda}=0.685$, $\sigma_8=0.829$, and $n_s=0.9655$) and the transfer function from \cite{Eisenstein1998}. We then selected a dark matter halo with mass ${\rm M_{\rm vir}=10^{11.8}M_{\odot}}$ at $z=6$ to resimulate at higher resolution.  Particles in and around this halo were traced back to the initial conditions at $z=150$ and regenerated at higher resolution giving an effective resolution of $4096^3$ particles in the Lagrange region of the halo.  The process of tracing particles back to $z=150$ and generating new initial conditions was repeated until the region surrounding the LBG was uncontaminated by low resolution dark matter particles out to twice the virial radius of the halo.  The final set of initial conditions were then regenerated to include gas, which is assumed to be initially neutral and composed of 76\% H and 24\% He by mass.  The dark matter particle mass in this final set of initial conditions is $4\times10^4{\rm M_{\odot}}h^{-1}$.  If we assume that a halo can be resolved by a minimum of 300 particles, the lowest mass resolved halo in our simulation has a mass of $1.2\times10^7{\rm M_{\odot}}h^{-1}$, well below the atomic cooling threshold mass.  The simulations have been run to $z=9.2$.

\subsection{Gravity, Hydrodynamics, Radiative Transfer, and Non-Equilibrium Chemistry}
We model gravity, hydrodynamics, radiative transfer, and non-equilibrium chemistry using the version of {\small RAMSES} presented in \cite{Katz2017,Kimm2017}.  Gravity is modelled using a multigrid scheme to solve the Poisson equation on the adaptive grid \citep{Guillet2011}.  Dark matter and star particles are projected onto the grid using cloud-in-cell interpolation.  Hydrodynamics is solved using the MUSCL-Hancock scheme with an HLLC Riemann solver \citep{Toro1994} and a MinMod slope-limiter.  In the hydrodynamic equations, we assume that the gas is ideal and monatomic and set $\gamma=5/3$ to close the relation between gas pressure and internal energy.

Radiative transfer (RT) is solved using a first-order moment method and we employ the M1 closure for the Eddington tensor \citep{Levermore1984} along with the Global-Lax-Friedrich intercell flux function.  Because the speed of light is much faster than the typical sound speed of the gas, modelling the radiation using the full speed of light would naturally lead to a simulation time step that is $\sim100-1000$ times smaller than a counterpart simulation without radiation, making the simulation computationally intractable.  For this reason, we artificially reduce the speed of light to $0.01c$, where $c$ is the speed of light, which is justified in high density regions \citep{Rosdahl2013}.  Furthermore, we apply an RT-subcyling method on each AMR level assuming Dirichlet boundary conditions at the fine-coarse interfaces \citep{Commercon2014,Rosdahl2018}.  We use up to a maximum of 500 RT subcycles per hydrodynamic time step on each AMR level.  Finally, we assume the on-the-spot-approximation such that all recombination radiation is absorbed locally.

\begin{table}
\centering
\begin{tabular}{@{}lcc@{}}
\hline
Group Name & $E_{\rm min}$ (eV) & $E_{\rm max}$ (eV) \\
\hline
Infrared & 0.1 & 1.0 \\ 
Optical & 1.0 & 5.6 \\
Habing & 5.6 & 11.2 \\
Lyman-Werner & 11.2 & 13.6 \\
HI-ionising &  13.6 & 15.2 \\
H$_2$-ionising & 15.2 & 24.59 \\
HeI-ionising &  24.59 & 54.42 \\
HeII-ionising & 54.42 & $\infty$ \\
\hline
\end{tabular}
\caption{Photon energy bins used in the simulation.  $E_{\rm min}$ and $E_{\rm max}$ represent the minimum and maximum photon energies used to define the edges of the bin, respectively. }
\label{photon_bin}
\end{table}

The radiation is split into eight energy bins: infrared, optical, Habing, Lyman-Werner, HI-ionising, H$_2$-ionising, HeI-ionising, and HeII-ionising, as listed in Table~\ref{photon_bin}, and is coupled to the gas via photoionisation, photo-heating, and radiation pressure.  Photoionisation and photo-heating are modelled using a seven-species non-equilibrium chemistry module that tracks HI, HII, e$^-$, H$_2$, HeI, HeII, and HeIII.  Details regarding the implementation of H and He non-equilibrium chemistry and the coupling to the radiation are described in \cite{Rosdahl2013} while the H$_2$ implementation is discussed in \cite{Katz2017}.  Radiation pressure is modelled both in the UV and optical (by single absorption) and in the IR via multiple scatterings on dust.  A description of the radiation pressure module can be found in \citet{Rosdahl2015} and we have used the {\small rt\_isoPress} implementation to alleviate the under-estimation of momentum transfer in the cells where the Stromgren sphere is unresolved.  We assume a mean dust opacity of 10$f_{\rm d,m}Z/Z_{\odot}$cm$^2$g$^{-1}$ in the IR radiation bin and 10$^3$$f_{\rm d,m}Z/Z_{\odot}$cm$^2$g$^{-1}$ in all other radiation bins (where $f_{\rm d,m}$ is the dust-to-metal ratio), consistent with \citet{Rosdahl2015b}.  Furthermore, if the temperature of a gas cell is $>10^5$K, we assume that the dust has been destroyed.  The formation and destruction mechanisms of dust are not explicitly modelled by our simulation, nor is dust considered a separate fluid.  Rather, we take a more simplistic approximation that the dust mass scales with the metallicity of the gas cell.  We employ the metallicity dependent dust-to-metal ratios from \cite{Remy2014} as was used by \cite{Kimm2018}.

Non-equilibrium cooling from collisional ionisations, recombinations, collisional excitation, bremsstrahlung, Compton cooling (and heating), and dielectronic recombination are computed for H and He and their ions (see Appendix~E of \citealt{Rosdahl2013}).  Furthermore, we use the H$_2$ cooling rates from \cite{Hollenbach1979}.  Metal-line cooling is also included in the simulation.  At $T>10^4$K, we employ cooling tables that are dependent on temperature and density and have been computed with {\small CLOUDY} \citep{Ferland1998}.  The cooling rate is assumed to scale linearly with metallicity.  At $T\leq10^4$K, we calculate the metal-line cooling rate from the fitting function of \cite{Rosen1995}. 

\subsection{Star Formation and Stellar Feedback}
Star particles are the only source of radiation in the simulation and we use the BPASSv2.0 models \citep{Stanway2016,Eldridge2008} to calculate the spectral energy distribution (SED). During each simulation time step, we use this SED to calculate the number of ionising photons that are dumped into the host cell of each star particle based on its age and metallicity.  Each star particle is assumed to be a single stellar population that has a stellar initial mass function with a maximum mass of 300M$_{\odot}$ and an IMF power-law slope of $-1.30$ between 0.1 to 0.5M$_{\odot}$ and $-2.35$ from 0.5 to 300M$_{\odot}$. The upper mass limit for our IMF is motivated by observations of local star clusters that show evidence for stars with $M>150{\rm M_{\odot}}$ \citep{Crowther2010} and the slope of the IMF is consistent with \cite{Kroupa2001}.  Note that our results may be sensitive to our choice of IMF parameters because the number of ionising photons can change considerably for different models for IMF, binary fraction, and stellar rotation (see \citealt{Stanway2016}).

Star particles are formed according to a thermo-turbulent recipe and this only occurs inside the zoom-in region and on the maximum AMR refinement level.  We require that a gas cell has $\rho_{\rm gas}>100{\rm cm^{-3}}$, the gas kinematics are locally convergent, the cell is a local density maximum, and the turbulent Jeans length (e.g. see Equation~8 of \citealt{Kimm2017} or Equation~2 in \citealt{Rosdahl2018}) is unresolved.  Note that we apply the method of  \cite{Rosdahl2018} and subtract rotational velocities and the symmetric divergence from the turbulent velocity dispersion when computing the turbulent Jeans length, in contrast to \cite{Kimm2017}.  If a cell satisfies all conditions, star formation is modelled using a Schmidt law \citep{Schmidt1959} such that:
\begin{equation}
\dot{\rho}_{*}=\epsilon_*\rho_{\rm gas}/t_{\rm ff},
\end{equation}
where $\dot{\rho}_{*}$ is the star formation rate density, $t_{\rm ff}$ is the free-fall time of the gas, and $\epsilon_*$ is the star formation efficiency.  $\epsilon_*$ is calculated based on the turbulent properties of the gas (based on \citealt{Federrath2012}) following Equation~2 of \cite{Kimm2017}.  The number of newly formed star particles in each cell for each simulation time step is drawn stochastically from a Poisson distribution with a minimum stellar mass of 1000M$_{\odot}$.

In the 50~Myr after a star particle is formed, it can undergo supernova (SN) explosions.  These are randomly drawn from a delay-time distribution \citep{Kimm2015}.  We employ the mechanical feedback model described in  \cite{Kimm2015,Kimm2017,Rosdahl2018} and the amount of momentum injected into the gas depends on the phase of the SN that is resolved by the simulation as to capture the final momentum of the snowplow phase.  The equivalent of $10^{51}$ergs is injected into the gas for each SN.  The maximum momentum that we inject is boosted according to \cite{Geen2015} to account for unresolved HII regions \citep{Kimm2017}.  For each massive star that explodes, 20\% of the mass is recycled back into the gas.  This gas is metal enriched assuming a metallicity of 0.075.  Following \cite{Rosdahl2018}, we have calibrated the SN feedback in order to reproduce the high-redshift stellar mass-halo mass relation from abundance matching \citep{Behroozi2013}.  For our simulation, this requires assuming that the mean SN progenitor mass is 5M$_{\odot}$ which leads to 4$\times$ more SN on average compared to a standard Kroupa IMF \citep{Kroupa2001}.  While not ideal, this results in galaxies that fall nicely on the stellar mass-halo mass relation (see Figure~1 of \citealt{Katz2018}) and produces a UV luminosity function consistent with observations for a similar set of simulations at approximately the same resolution \citep{Rosdahl2018}.  While we cannot be certain that even our calibrated simulations have the correct stellar mass-halo mass relation as we are using a high-redshift extrapolation and there is intrinsically a lot of scatter, we have made an effort to calibrate on one of the best available predictions as any significant offset from the stellar mass-halo mass relation may lead to large systematic offsets in the SFR-line luminosity relations. In summary, our star formation and stellar feedback models are based on the work of \cite{Rosdahl2018} which were chosen to reproduce both a reasonable reionization history and UV luminosity function.

\subsection{Refinement}
Grid refinement occurs on-the-fly in the simulation.  When a cell contains either eight dark matter particles or the gas mass of the cell is $>8\frac{\Omega_{\rm b}}{\Omega_{\rm DM}}{\rm m_{DM}}$, where ${\rm m_{DM}}$ is the mass of a dark matter particle, it is flagged for refinement.  Furthermore, cells are flagged if the cell width is more than a quarter of the local Jeans length.  A cell is refined by splitting it into 8 children cells.  Our refinement strategy aims to maintain a constant physical resolution of 13.6 parsecs in the simulation and this is achieved by releasing new levels of refinement at fixed increments of the cosmological scale factor.  

\subsection{Halo Finder}
In order to extract haloes from the simulation, we use the AMIGA halo finder \citep[AHF,][]{Gill2004,Knollmann2009}.  This is applied in post-processing to nine different simulation snapshots at $z=12.0,$ 11.5, 11.0, 10.5, 10.0, 9.8, 9.6, 9.4, and 9.2.  We define the virial radius of a galaxy to be that which contains a mean density of stars, gas and dark matter equal to $\Delta\rho_{\rm crit}$, where $\rho_{\rm crit}$ is the critical density of the Universe and $\Delta$ is the over-density that would allow for spherical collapse against an expanding cosmological background.  The value of $\Delta$ is set for each redshift for our cosmology and at the redshifts we are interested in, $\Delta\sim200$.  In our analysis, we only study haloes that are free from contamination by low resolution dark matter particles and consist of at least 300 high resolution dark matter particles.  At $z=10$, there are more than 1,000 uncontaminated haloes in our simulation with a mass $>1.2\times10^7{\rm M_{\odot}}h^{-1}$.

\subsection{Estimating Nebular and IR Line-Luminosities}
We estimate nebular and IR line luminosities using the spectral synthesis code {\small CLOUDY} \citep{Ferland2017}.  For each simulation cell, we know the temperature, density, metallicity, and local radiation field.  For each cell in the simulation, we would ideally set up a {\small CLOUDY} simulation, matching the properties of the gas and radiation field to calculate the line emission; however, due to the sheer number of cells in the simulation, this is computational impractical.  Furthermore, we cannot simply create a lookup table because we describe the cell by nine parameters ($T$, $\rho$, $Z$, and six for the radiation field), and sparsely sampling this space using ten grid points per parameter would also result in too many models to be computationally feasible.  Therefore, we have developed a different approach.  

We begin by extracting the properties of $\sim850,000$ cells from the central regions of our most massive galaxy at $z=10$.  These properties include temperature, density, metallicity, the flux in the six radiation bins at $E>5.6$eV\footnote{Note that we neglect the IR and Optical bins as they are extremely coarsely sampled in our simulations.}, and the length of the cell.  We then set up a {\small CLOUDY} model for each of these $\sim850,000$ simulation cells using a gas slab in an open geometry such that the slab has the same density, constant temperature, metallicity, depth, radiation flux consistent with that of our simulation cell, and an isotropic background from the CMB at $z=10$.  We assume that the abundance ratios of individual elements are consistent with Solar \citep{Grevesse2010} and we allow the models to iterate until convergence.  The fluxes for each line are then taken as the emergent fluxes output by {\small CLOUDY}.  

In order to calculate the line luminosities of all other cells in the simulation, both in the $z=10$ snapshot and in others, we employ a random forest machine learning algorithm \citep{Ho1995,Breiman2001,Geurts2006}.  The Random Forest (RF) algorithm is an example of an ensemble machine learning method in that it employs large numbers of estimators simultaneously. The RF draws on the outputs of the estimators to reach an overall prediction. For regression problems (as is the case here), we have chosen the output to be the average of the outputs of each estimator.

The estimators used in a RF are decision trees that partition data into a several subsets. Each datapoint contains the values for each ``feature" of our data. In our case, the features are temperature, density, metallicity, the flux in the six radiation bins at $E>5.6$eV, and the length of the cell. The decision tree consists of nodes where binary decisions about a particular feature of each data point arriving at the node occur. We can encode the behaviour of each node by specifying the feature of interest to the node, and the threshold value of that feature for the partition of the data.  We grow our RF using 80\% of our total data. We call this subset the training set. The final 20\% is reserved for testing the accuracy of the RF after training.  

We use the ensemble random forest regressor implementation in {\small scikit-learn}\footnote{http://scikit-learn.org/stable/} without setting a maximum depth of the trees so that they can become as arbitrarily deep as needed for the most accurate prediction.  We vary the number of trees in the forest from 50 to 500 and use cross-validation to settle on 100 as the optimal number of trees for the algorithm.  

\begin{figure}
\centerline{\includegraphics[scale=1]{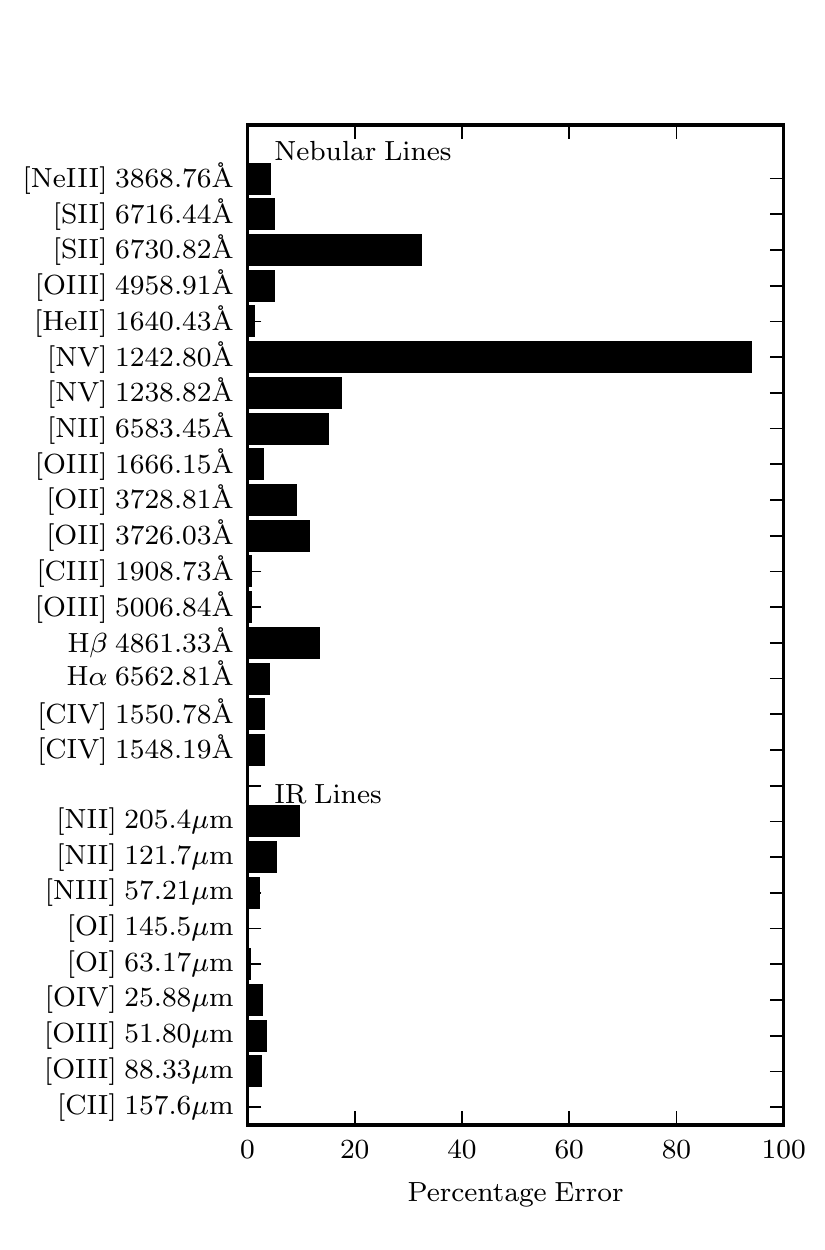}}
\caption{Estimated fractional percentage error ($|L_{\rm RF}-L_{\rm CLOUDY}|/L_{\rm CLOUDY}$) in the total luminosities of each line that we model based on the performance of the trained random forests on the ``test set".  Typical accuracies for the brighter lines in our simulation computed with the random forest are within $\sim5\%$ of that measured with {\small CLOUDY}.  For the ``test set", all except two lines have errors less than $\sim20\%$, while our most inaccurate line, [NV]~$1242.80\angstrom$ is accurate within a factor of two.  }
\label{lineerror}
\end{figure}

For each line we aim to predict, we train a separate RF\footnote{This was done so we could find the optimal number of trees for each line.}.  Training the RFs is usually complete in a few minutes and execution on an individual set of parameters is nearly instantaneous.  The idea behind this is that we need to be able to obtain the results that {\small CLOUDY} would give, without having to run {\small CLOUDY} due to its computational expense.  In our case, executing the RF on millions of cells takes only a matter of seconds whereas, assuming the optimistic case where each {\small CLOUDY} model takes 8s to run.  Executing {\small CLOUDY} models across a simulation that has 100,000,000 cells across 10 simulation outputs would require 2.2~Million CPU hours.  In our current work, we make predictions for $\sim67$ million cells, hence we argue that our machine learning method is optimally suited for the problem at hand.

\begin{table}
\centering
\begin{tabular}{@{}lclc@{}}
\hline
Line & Wavelength & Line & Wavelength  \\
\hline
{\bf Infrared} & & &\\
{[OIV]} & 25.8832$\mu$m  & {[OIII]} & 51.8004$\mu$m  \\

{[NIII]} & 57.3238$\mu$m  & {[OI]} &  63.1679$\mu$m  \\

{[OIII]} & 88.3323$\mu$m & {[NII]} & 121.767$\mu$m  \\

{[OI]} & 145.495$\mu$m  & {[CII]} & $157.636\mu$m  \\ 

{[NII]} & 205.224$\mu$m & & \\
\hline
{\bf Nebular} & & & \\
{[NV]} & 1238.82$\angstrom$ & {[NV]} & 1242.80$\angstrom$ \\

{[CIV]} & 1548.19$\angstrom$ & {[CIV]} & 1550.78$\angstrom$ \\

{[HeII]} & 1640.43$\angstrom$ & {[OIII]} & 1666.15$\angstrom$  \\

CIII] & 1908.73$\angstrom$  & {[OII]} & 3726.03$\angstrom$   \\

{[OII]} & 3728.81$\angstrom$  & {[NeIII]} & 3868.76$\angstrom$  \\

H$\beta$ & 4861.33$\angstrom$ & {[OIII]} & 4958.91$\angstrom$  \\

{[OIII]} & 5006.84$\angstrom$  & H$\alpha$ & 6562.81$\angstrom$  \\

{[SiII]} & 6716.44$\angstrom$  & {[SiII]} & 6730.82$\angstrom$  \\

\hline
\end{tabular}
\caption{List of lines and wavelengths that we study in this work.}
\label{linelist}
\end{table}

\begin{figure*}
\centerline{\includegraphics[scale=1]{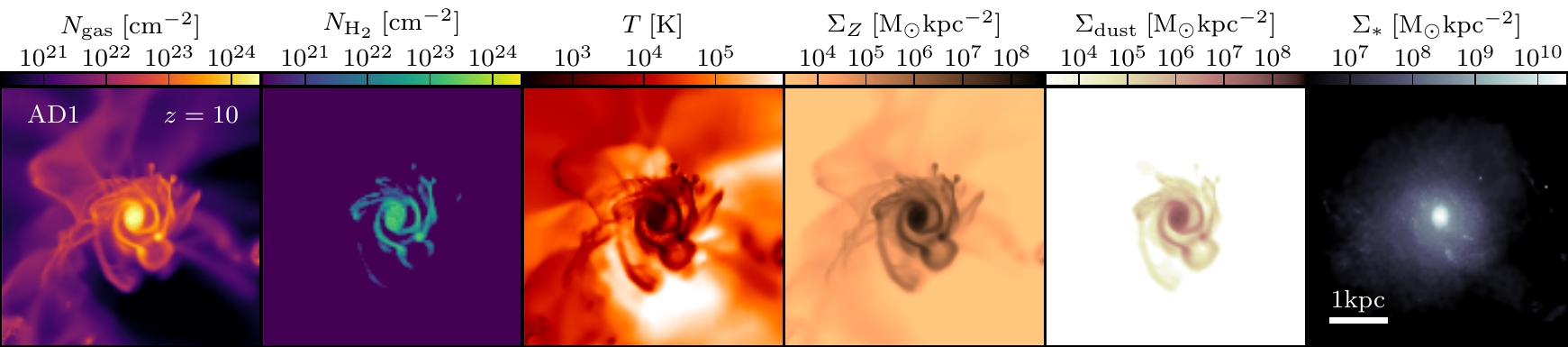}}
\centerline{\includegraphics[scale=1]{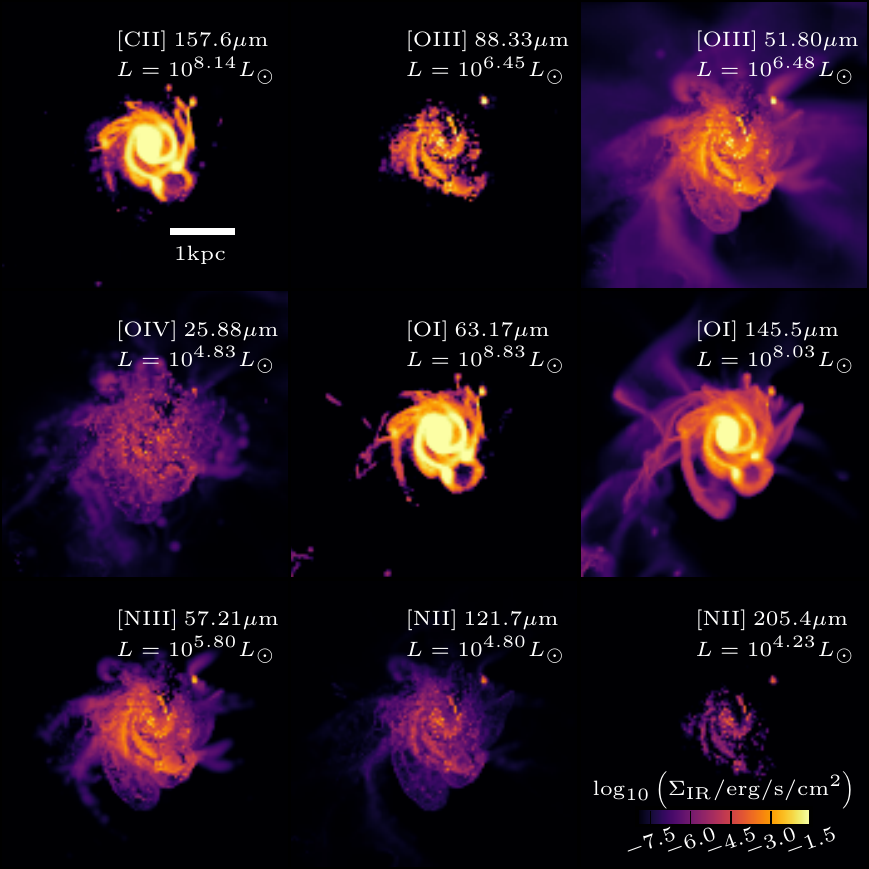}\includegraphics[scale=1]{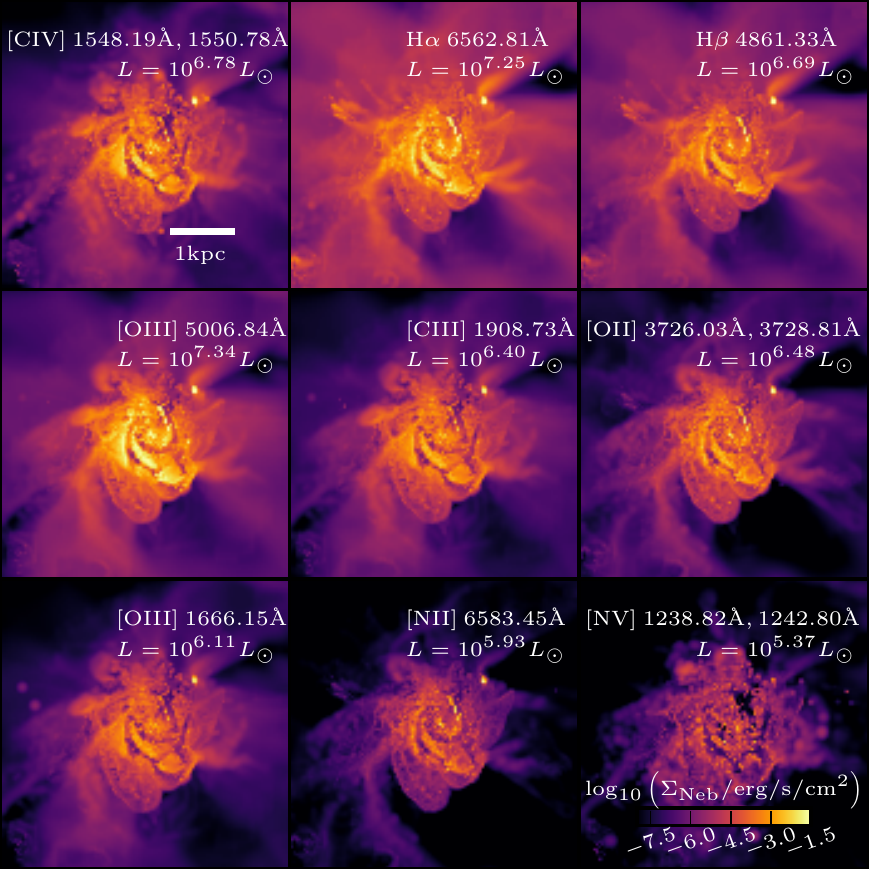}}
\caption{Face-on view of AD1 at $z=10$.  The top row shows maps of total gas column density, H$_2$ column density, density weighted temperature, metal surface mass density, dust surface mass density, and stellar mass surface density in a 5~kpc cube surrounding the galaxy.  The bottom three rows show the surface brightness of different infrared and nebular lines.  The galaxy appears as a well formed disk with ordered rotation.  }
\label{ALMA_FO}
\end{figure*}

The major open question is the accuracy of the method.  As stated before, we have reserved 20\% of the $\sim850,000$ cells as a ``test set" to test the accuracy of the method.  This is designed to determine how well the trained algorithm generalises, or in other words, how it performs on data that it has never seen before.  In Figure~\ref{lineerror}, we show the fractional percentage error ($|L_{\rm RF}-L_{\rm CLOUDY}|/L_{\rm CLOUDY}$, where $L_{\rm RF}$ is the luminosity predicted by the random forest and $L_{\rm CLOUDY}$ is the actual {\small CLOUDY} luminosity) in predicting the total line luminosity of the ``test set", for each line listed in Table~\ref{linelist}.  In general, we can predict the total luminosity to within $\sim10\%$ accuracy for nearly all our lines, which is well within the uncertainties and differences one would obtain by changing certain model parameters such as the stellar SED, cloud geometry, stellar feedback model, etc.  For particularly bright lines, such as [CII]~157.6$\mu$m, we can predict the galaxy luminosity to better than $1\%$ accuracy.  Indeed our model is more accurate for some lines than others.  In particular, we find large inaccuracies for [NV]~1242.80$\angstrom$.  This line is very weak in our simulated galaxies because it requires very highly ionised gas (and is often associated with AGN) and is thus more difficult to predict.  However, for our purposes, these errors are not particularly large because in the absolute worst possible scenario, our line predictions are still within a factor of two of those predicted by {\small CLOUDY}. 

Having demonstrated that our new method is accurate enough to apply to our simulations, we now apply the RF to all cells within the halos of interest in order to calculate the luminosity of each halo in each line.

\begin{figure*}
\centerline{\includegraphics[scale=1]{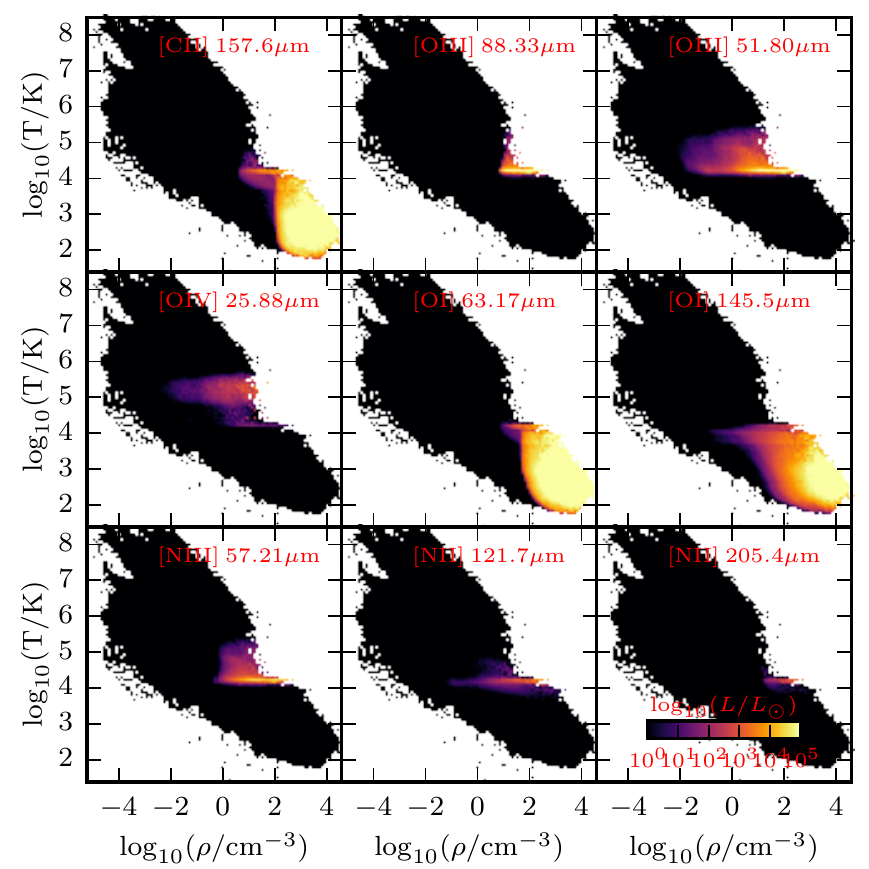}\includegraphics[scale=1]{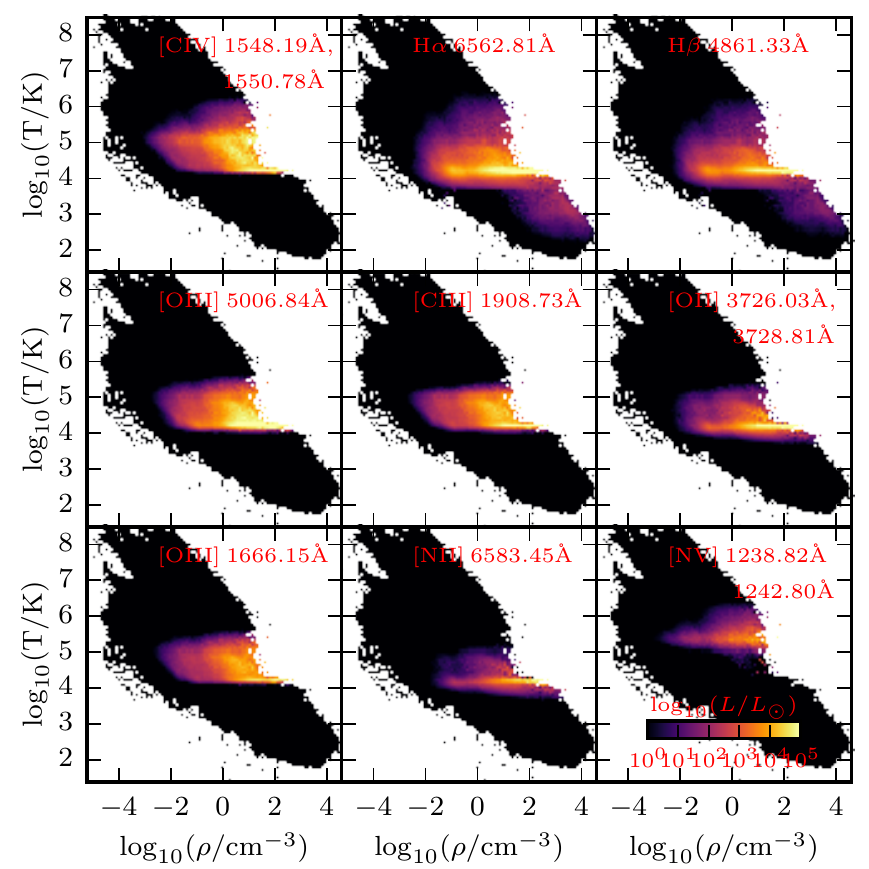}}
\caption{Phase-space diagrams of gas density versus temperature for all cells in a sphere of radius 5kpc around the centre of AD1 at $z=10$.  The 2D histograms have been weighted based on the luminosity of an individual emission line in each pixel, while the black silhouette represents the regions of this diagram mapped out by all cells in this sphere.  There are 100 log-spaced pixels on each axis with widths of 0.1~dex in $\rho$ and 0.07~dex in $T$.  Each panel shows the gas conditions where the majority of the luminosity in each line is being emitted.}
\label{ALMA_ps}
\end{figure*}

\subsection{Dust Attenuation Calculation}
Certain lines are subject to dust attenuation in the ISM.  To calculate this, we use a similar method as presented in \cite{Katz2018}.  For each galaxy, we interpolate the dust properties of the central 5kpc onto a $500\times500\times500$ cube such that each cell in the cube has a length of $\sim10$pc which is approximately the maximum physical resolution of the simulation.  We assume that the dust content scales with the metallicity of the simulation cell following \cite{Remy2014,Kimm2018} and in all cells with $T>10^5$K, we assume that the dust has been destroyed.  We focus only on the central 5kpc for each galaxy as this region contains most of the emission and allows us to create a dust cube with similar resolution as the maximum of the simulation (larger regions would be much more computationally and memory intensive).  We calculate the optical depth of dust using the $R_V=3.1$ Milky Way dust grain model of \cite{Weingartner2001} to attenuate the SED.  We then attenuate the lines according to the optical depth to dust along the line of sight for the given wavelength.  Note that we almost always use the intrinsic line luminosity when comparing to observations (as the observations are generally corrected for reddening) and state specifically when we use the dust attenuated versions.

\begin{table*}
\centering
\begin{tabular}{@{}lccccccccc@{}}
\hline
Property & $z=12.0$ & $z=11.5$ & $z=11.0$ & $z=10.5$ & $z=10.0$ & $z=9.8$ & $z=9.6$ & $z=9.4$ & $z=9.2$\\
\hline
\hline
{\bf AD1} & & & & & & & & &  \\
Halo Mass & 10.38 & 10.46 & 10.54 & 10.62 & 10.83 & 10.91 & 10.98 & 11.02 & 11.06  \\
Stellar Mass & 8.76 & 8.89 & 9.04 & 9.13 & 9.37 & 9.41 & 9.49 & 9.54 & 9.58  \\
Gas Mass & 9.51 & 9.60 & 9.68 & 9.75 & 10.02 & 10.08 & 10.15 & 10.19 & 10.22  \\
SFR & 6.21 & 13.99 & 18.51 & 8.23 & 16.51 & 20.37 & 25.99 & 29.35 & 24.10  \\
Metallicity & 1.87e-3 & 1.96e-3 & 2.15e-3 & 2.71e-3 & 2.57e-3 & 2.49e-3 & 2.45e-3 & 2.51e-3 & 2.61e-3  \\
H$_2$ Mass & 8.26 & 8.35 & 8.41 & 8.67 & 8.90 & 8.92 & 8.96 & 8.99 & 9.04  \\
\hline
{\bf AD2} & & & & & & & & &  \\
Halo Mass & 9.96 & 10.01 & 10.08 & 10.13 & 10.16 & 10.18 & 10.20 & 10.21 & 10.22 \\
Stellar Mass & 7.83 & 7.94 & 7.96 & 8.01 & 8.10 & 8.16 & 8.34 & 8.37 & 8.41  \\
Gas Mass & 8.99 & 9.07 & 9.21 & 9.31 & 9.38 & 9.43 & 9.45 & 9.50 & 9.52  \\ 
SFR & 0.78 & 0.47 & 0.24 & 0.53 & 1.28 & 1.39 & 5.34 & 1.28 & 1.96  \\
Metallicity & 5.98e-4 & 7.46e-4 & 6.64e-4 & 5.75e-4 & 5.75e-4 & 5.95e-4 & 7.05e-4 & 8.74e-4 & 9.81e-4  \\
H$_2$ Mass & 7.28 & 7.08 & 7.18 & 7.55 & 7.42 & 8.01 & 7.89 & 7.87 & 8.03  \\
\hline
{\bf AD3} & & & & & & & & &  \\
Halo Mass & 9.71 & 9.77 & 9.82 & 9.87 & 9.92 & 9.93 & 9.95 & 9.98 & 10.03  \\
Stellar Mass & 8.11 & 8.25 & 8.31 & 8.40 & 8.48 & 8.49 & 8.51 & 8.54 & 8.58  \\
Gas Mass & 8.91 & 8.95 & 9.00 & 9.05 & 9.11 & 9.13 & 9.15 & 9.20 & 9.27  \\
SFR & 2.91 & 2.94 & 1.20 & 2.43 & 2.18 & 1.12 & 1.42 & 1.66 & 2.06  \\
Metallicity & 1.28e-3 & 2.12e-3 & 2.69e-3 & 2.83e-3 & 3.00e-3 & 3.09e-3 & 3.09e-3 & 2.93e-03 & 2.59e-3  \\
H$_2$ Mass & 7.41 & 7.63 & 8.09 & 8.08 & 8.10 & 8.30 & 8.31 & 8.32 & 8.30 \\
\hline
\end{tabular}
\caption{Properties of AD1, AD2, and AD3 as a function of redshift.  All masses are quoted in $\log_{10}({\rm M/M_{\odot}})$.  Metallicities represent the mean mass-weighted gas-phase metallicity, while the SFRs are in units of ${\rm M_{\odot}/yr}$ averaged over 10Myr.}
\label{hprops}
\end{table*}

\section{Results - General Properties}
\label{results_GP}
We focus the majority of our analysis on the evolution of the three most massive systems in the simulation, AD1, AD2, and AD3, which have halo masses of $10^{11.1},\ 10^{10.2}$, and $10^{10.1}{\rm M_{\odot}}$ and SFRs of 33.43, 4.92, and 1.52 M$_{\odot}$~yr$^{-1}$ at $z=9.0$, respectively.  The properties of these three galaxies at different redshifts can be found in Table~\ref{hprops}.  Each one of these systems exhibits a different star formation history, with AD1 and AD2 both turning on $\sim100$Myr after the Big Bang while AD3 forms its first star $\sim100$Myr after AD1 and AD2.  By $z=9$, AD1, AD2, and AD3 have total stellar masses of $10^{9.63},\ 10^{8.50},$ and $10^{8.61}{\rm M_{\odot}}$, respectively, such that they fall nicely on the extrapolated stellar mass-halo mass relation predicted from abundance matching at $z=9$ \citep{Behroozi2013}.  The diversity of SFHs and cumulative stellar mass growths as well as the stellar mass-halo mass relation for galaxies in our simulation can be found in Figure~3 of \cite{Katz2018}.  Despite having very similar stellar masses, halo masses, and SFRs, AD2 and AD3 exhibit completely different morphologies with AD2 being considerably more diffuse.  Similarly, AD2 has a mean gas phase metallicity that is more than a factor of two smaller than AD3 by $z=9$.  A complete description of the basic evolutionary properties and the continuum SEDs of these three galaxies can be found in \cite{Katz2018} and in what follows, we focus primarily on the emission line properties of these three systems.

\begin{figure*}
\centerline{\includegraphics[scale=1,trim={0 0.5cm 0 0.18cm},clip]{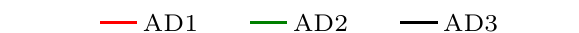}}
\centerline{\includegraphics[scale=1]{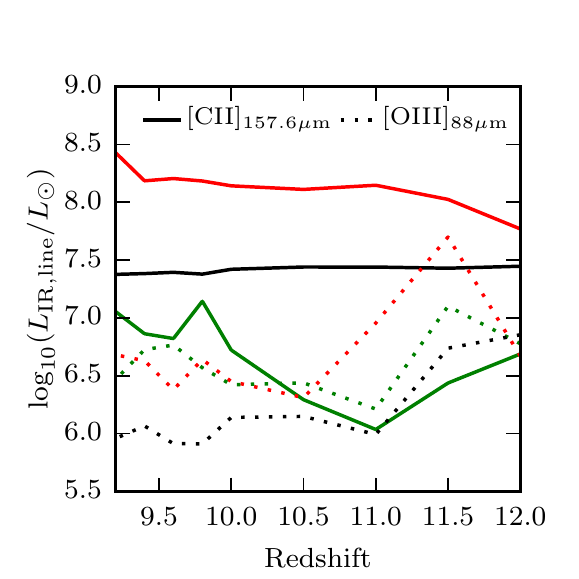}\includegraphics[scale=1]{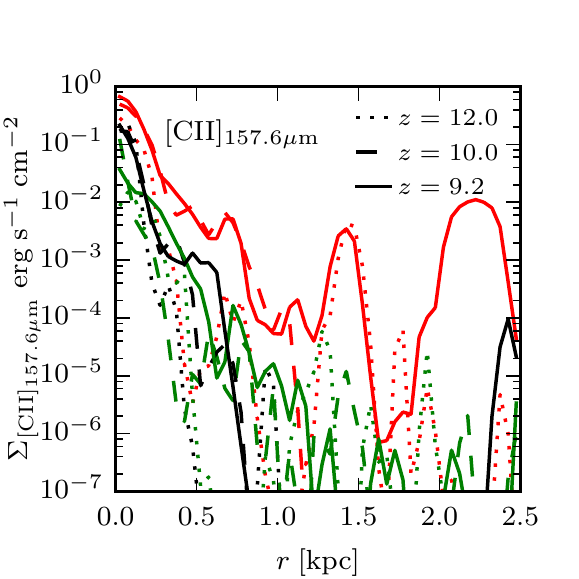}\includegraphics[scale=1]{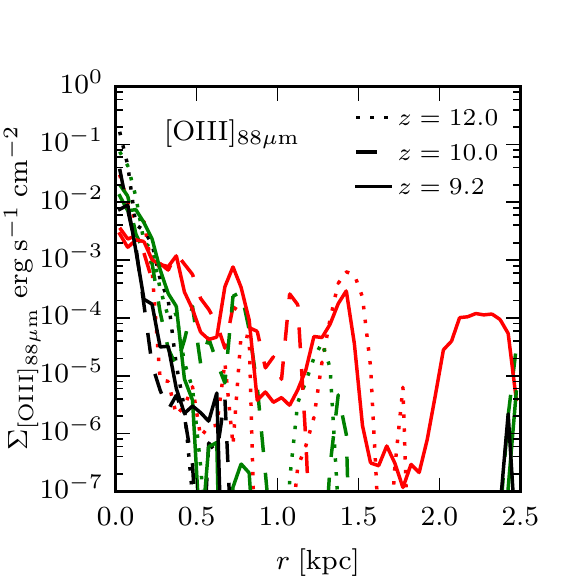}}
\caption{(Left) Evolution of the [CII]~157.6${\rm \mu m}$ (solid) and [OIII]~88${\rm \mu m}$ (dotted) luminosities as a function of redshift.  The results for AD1, AD2, and AD3 are shown in red, green, and black, respectively.  (Centre) Face-on surface brightness profiles of [CII]~157.6${\rm \mu m}$ at $z=12.0,\ 10.0,\ \&\ 9.2$ (dotted, dashed, solid) for AD1, AD2, and AD3. (Right) Face-on surface brightness profiles of [OIII]~88${\rm \mu m}$  at $z=12.0,\ 10.0,\ \&\ 9.2$ for AD1, AD2, and AD3. }
\label{IR_line_redshift}
\end{figure*}

In Figure~\ref{ALMA_FO}, we show an illustrative example of AD1 at $z=10$.  The top row shows face-on maps of total gas column density, H$_2$ column density, density weighted temperature, metal surface mass density, dust surface mass density, and stellar mass surface density in a 5kpc cube surrounding the galaxy.  Similarly, face-on images for nine IR lines and twelve of the nebular lines listed in Table~\ref{linelist} (doublet lines have been combined into single images) can be found in the bottom three rows.  Each one of these lines has a morphology depending on the phase of the ISM from which it originates, the critical density, and the metal enrichment of the gas.  The most commonly observed IR lines, [CII]~157.6$\mu$m and [OIII]~88.33$\mu$m, have very different morphologies.  [CII]~157.6$\mu$m is very well correlated with the molecular content, which is consistent with many other works that have modelled these systems \citep{Katz2017,P17a,P17b,Olsen2017,Vallini2015}, while [OIII]~88.33$\mu$m is picking out individual star forming regions (see also \citealt{Katz2017,Moriwaki2018}).  In these images, we are showing the emergent emission rather than the intrinsic emission as in practice, these lines are observed against and modulated by the background CMB \citep{Chatzikos2013,Lagache2018}.  This tends to have the effect of diminishing the emission from the lower density regions of the galaxy.  Although they are not affected by the CMB, the nebular lines are affected by the presence of dust and each nebular line image that we show in Figure~\ref{ALMA_FO} takes into account the effect of dust attenuation.  This is calculated by measuring the dust optical depth to each cell in the simulation along the line-of-sight for each specific wavelength and reducing the emission from that cell accordingly.  Note how in the centres of many of the images where emission is expected to be the brightest, dark dust lanes exist demonstrating this attenuation.  These are consistent with the location of dust shown in the fifth panel of the top row.  Dust attenuation is more pronounced for the rest-frame UV lines compared to the rest-frame optical lines. 

It is instructive to identify the exact regions of temperature-density phase-space where each emission line originates.  In Figure~\ref{ALMA_ps}, we highlight the regions of the phase-space within a 5kpc sphere surrounding AD1 at $z=10$ based on the contribution to the total luminosity of each line.  The left panel shows IR lines while the right panel shows nebular lines. This provides a consistency check that our modelling is appropriately resolving the individual phases of the ISM where each line is expected to originate.  Both [CII]~157.6$\mu$m and [OI]~67.13$\mu$m are brightest in the highest density neutral regions with $\rho>100{\rm cm^{-3}}$ at temperatures of $T<10^4$K.  These are the same regions where H$_2$ forms efficiently.  In contrast, [OIII]~88.33$\mu$m is most luminous in the intermediate density, higher temperature regions where higher ionisation states can be reached.  Finally, [OIV]~25.88$\mu$m, our highest ionisation IR line, is brightest in gas with $T>10^5$K which requires either extremely high energy photons or SN.  Turning to the nebular lines, H$\alpha$ and H$\beta$ probe the lower temperature, recombining gas while the [NV]~1241$\angstrom$ doublet, the highest ionisation potential line shown, is clearly probing gas at much higher temperatures.  Note that the H$\alpha$ and H$\beta$ from cold $<10^4$K gas occurs in partially ionised, unresolved HII regions.  By combining the information from these different lines, we can begin to understand the physical properties of the ISM observed in high-redshift galaxies and these plots are very instructive for interpreting our comparisons with observations.

\begin{figure*}
\centerline{\includegraphics[scale=1]{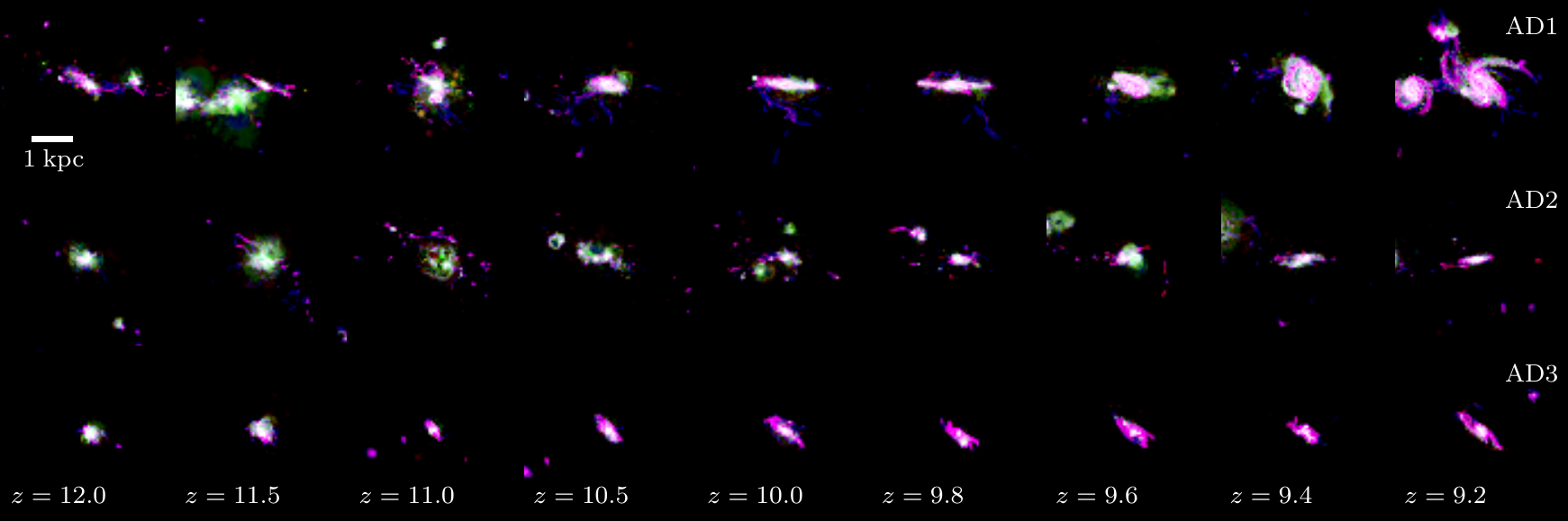}}
\caption{Three channel images showing the evolution and morphology of [CII]~157.6$\mu$m, [OIII]~88.33$\mu$m, and [OI]~67.13$\mu$m luminosity for the three galaxies, AD1, AD2, and AD3 at $12\geq z\geq9.2$.  Each image shows a projected distribution of emission lines for a 5~kpc cube surrounding the galaxy in an edge-on orientation where red represents [CII]~157.6$\mu$m, green represents [OIII]~88.33$\mu$m, and blue represents [OI]~67.13$\mu$m.  The purple regions indicate the locations of [CII]~157.6$\mu$m and [OI]~67.13$\mu$m as both emanate from the same high-density, cold, neutral gas, while the green regions trace the ionised warmer gas that has been affected by feedback.  }
\label{time_series}
\end{figure*}

Since IR lines have already been directly observed at high redshift, it is important to understand the time variability and morphology of the emission.  In the left panel of Figure~\ref{IR_line_redshift}, we show how the [CII]~157.6$\mu$m and [OIII]~88.33$\mu$m luminosities change as a function of redshift between $9.2\leq z\leq12.0$.  For AD1 and AD3, the [CII]~157.6$\mu$m emission is remarkably stable throughout this time period while the [OIII]~88.33$\mu$m emission can fluctuate by more than an order of magnitude.  In contrast, AD2 is more variable in both [CII]~157.6$\mu$m and [OIII]~88.33$\mu$m.  This may be related to the fact that the morphologies of AD1 and AD3 are completely different from that of AD2.  In Figure~\ref{time_series}, we show 3-channel RGB maps of the 5kpc region around the three galaxies in the same redshift interval, where each colour channel is representing the emission from a different line ([CII]~157.6$\mu$m in red, [OIII]~88.33$\mu$m in green, and [OI]~67.13$\mu$m in blue).  Note that the [OI]~67.13$\mu$m and [CII]~157.6$\mu$m emitting regions often mix to create purple.  The maps have been oriented with the principal angular momentum axis parallel to the page (edge-on).  AD1 and AD3 exhibit a well defined structure, and in some snapshots, even thin disks, while the morphology of AD2 is much more irregular and disordered.  However, the morphologies are very time variable.  AD1 only forms a disk at $z\sim10$ and it remains thin until $z\sim9.6$ where it begins to become disrupted.  This is due to the presence of satellite galaxies that are perturbing the central massive system and become visible in the $z=9.2$ snapshot.  In contrast, AD3 remains fairly stable between $z=10.5$ and $z=9.2$ as it seems to be evolving in relative isolation.   

The green regions in these images trace the sites of recent star formation and galactic winds where either high energy radiation or SNe are responsible for ionising the gas.  Of the three galaxies, AD2 seems most susceptible to SN feedback as bright green patches are seen throughout the evolution of AD2, indicating that the star formation and feedback are efficient at disrupting the system.  The constant disruption of AD2 may provide a link to the time variability of the [CII]~157.6$\mu$m and [OIII]~88.33$\mu$m emission.

In general, [CII]~157.6$\mu$m and [OIII]~88.33$\mu$m emission from these three galaxies is very centrally concentrated.  In the centre and right panels of Figure~\ref{IR_line_redshift}, we show the surface brightness profiles of [CII]~157.6$\mu$m and [OIII]~88.33$\mu$m at $z=12.0$, $z=10.0$, and $z=9.2$.  Apart from the $z=9.2$ and $z=12.0$ snapshot of AD1, almost all of the [CII]~157.6$\mu$m emission is concentrated in the central 1kpc and the surface brightness falls off by more than six orders of magnitude out to this radius.  At $z=9.2$, AD1 is undergoing a merger and the radial profile reflects the presence of the two satellite systems.  Our only example of an extended [CII]~157.6$\mu$m profile occurs when satellite systems are present.

\section{Results - Comparison with Observations}
\label{results_OC}
Having described some of the general properties of the high-redshift galaxy population in our simulation, in the following sections, we focus on directly comparing our simulations with individual properties of observed high-redshift galaxies.

\subsection{Infrared Lines}
As was described in Section~\ref{intro}, deep IR observations of high-redshift galaxies have revealed numerous interesting features such as spatial and spectral offsets between different far-infrared emission lines and the UV/Ly$\alpha$, deviations from the local [CII]-SFR relation, weaker far-infrared emission compared to similar low-redshift systems, and velocity gradients signifying rotation.  Our simulations are well suited to explain the underlying physics governing these properties. 

\subsubsection{ISM Kinematics with IR Lines}
Recently, \cite{Smit2018} detected velocity gradients in [CII]~157.6$\mu$m emission in two galaxies at $z\sim6.8$.  Three rather different explanations for the velocity gradients have been suggested. These include, assuming that the system is rotating, that one is witnessing the merger of two systems, or that some of the [CII]~157.6$\mu$m emission originates in an outflow.  If the dynamics of these galaxies are dominated by rotation, the internal structure has a similar ratio of rotation to turbulence as seen at $z\sim2$ \citep{Smit2018,Forster2009}.  However, it remains to be determined whether the systems are indeed rotating because other explanations remain viable.

In Figure~\ref{ALMA_FO}, signs of ordered rotation are indeed present for AD1.  This system clearly has a transient disc-like structure.  Our simulation is not unique in this regard, as other simulations have also found rotation despite using different feedback models \citep{P17a,P17b}.  Furthermore, from Figure~\ref{time_series}, both AD2 and AD3 also exhibit some evidence for ordered rotation.  The environments around these systems are highly dynamic and mergers occur relatively frequently at these redshifts, and the velocity structure of these systems thus changes rapidly.  Nevertheless, many of the images in Figure~\ref{time_series} show an object that appears to be rotating and hence, with high resolution spectroscopy, one would expect to observe velocity gradients for these type of galaxies. 

\begin{figure*}
\centerline{\includegraphics[scale=1]{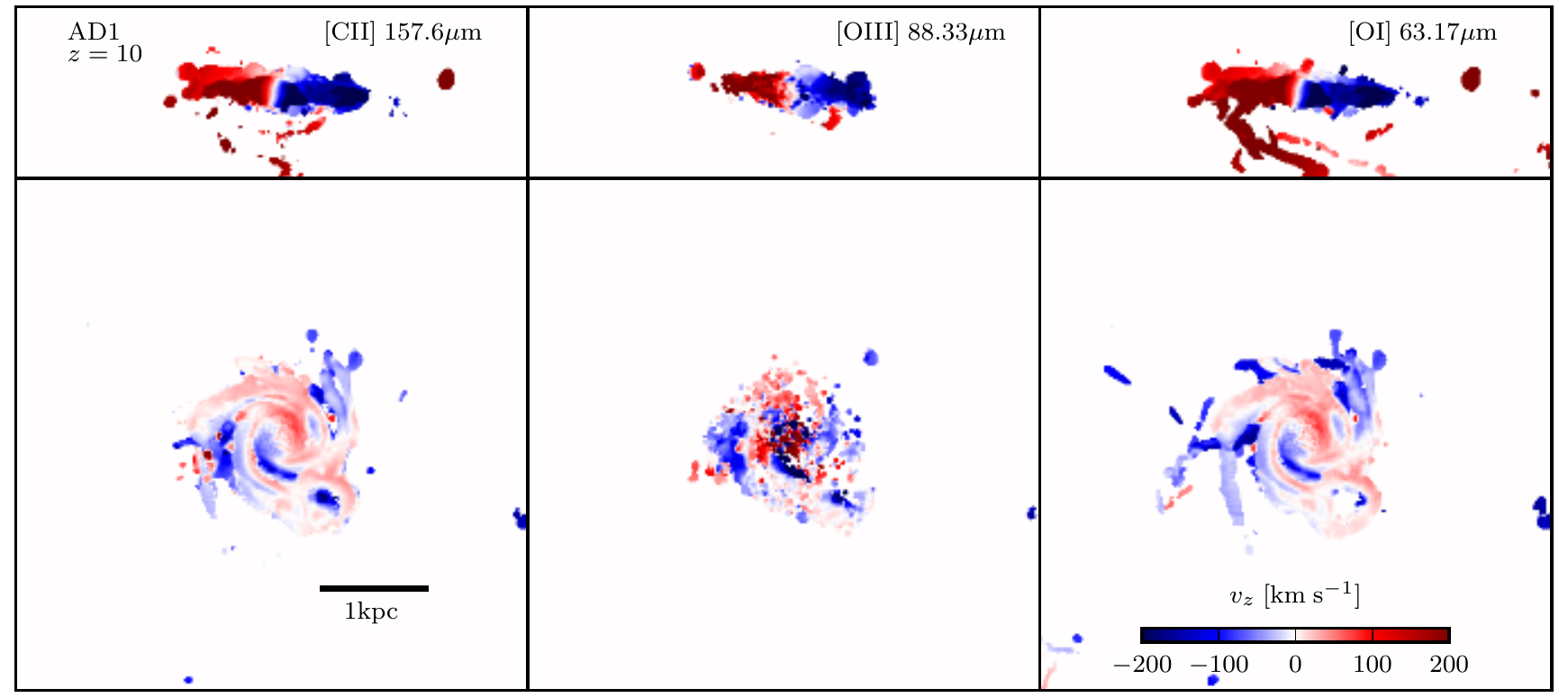}}
\caption{Velocity maps of AD1 weighted by [CII]~157.6$\mu$m (left), [OIII]~88.33$\mu$m (middle), or [OI]~63.17$\mu$m (right) surface brightness for the $z=10$ output.  We only show regions with a surface brightness $>10^{-6.5}$erg/s/cm$^2$. $v_z$ represents the velocity perpendicular to the page and thus a velocity of 0 km~s$^{-1}$ indicates a velocity that is tangential to the line-of-sight.  The top row shows the system edge-on while the bottom row shows the system face-on.  Clear velocity gradients are observed for the system in the edge-on orientation.}
\label{ALMA_vm}
\end{figure*}

In Figure~\ref{ALMA_vm}, we show velocity maps for the $z=10$ snapshot of AD1, both edge-on and face-on in a 5kpc box around the system, for [CII]~157.6$\mu$m, [OIII]~88.33$\mu$m, and [OI]~63.17$\mu$m.  The edge-on maps in the top row of Figure~\ref{ALMA_vm} show a strong velocity gradient on both sides of the disc that reach velocities of $\pm 200$km/s on either side.  

If the disks are viewed face-on, there are no velocity gradients, as expected.  The [OIII]~88.33$\mu$m image has a bright red spot in the central region which may be associated with gas outflowing from the centre of the disk.  Since real galaxies are unlikely to be viewed perfectly edge-on, we expect the observed velocity gradients to be significantly lower than 200km/s for this type of system.  \cite{Smit2018} observe velocity gradients of up to $\pm60$km/s on approximately the same scale as our images, perhaps suggesting lower host halo masses than that of our simulated galaxies AD1-3.  Note that the observations are at much lower spatial resolution than our simulations which can also decrease this signal.  Given the ubiquity of rotation among the massive galaxies in our simulation at some point during their evolution, we expect this to be observable at high-redshift and can confirm that ordered rotation is a plausible (but perhaps not the only) explanation for the \cite{Smit2018} objects.

\subsubsection{Spatial and Spectral Offsets Between IR Lines}

\begin{figure}
\centerline{\includegraphics[scale=1]{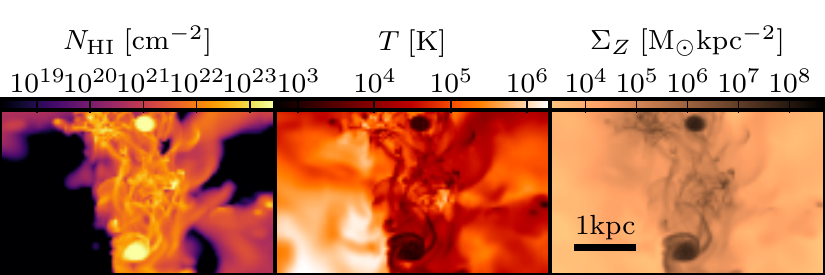}}
\centerline{\includegraphics[scale=1]{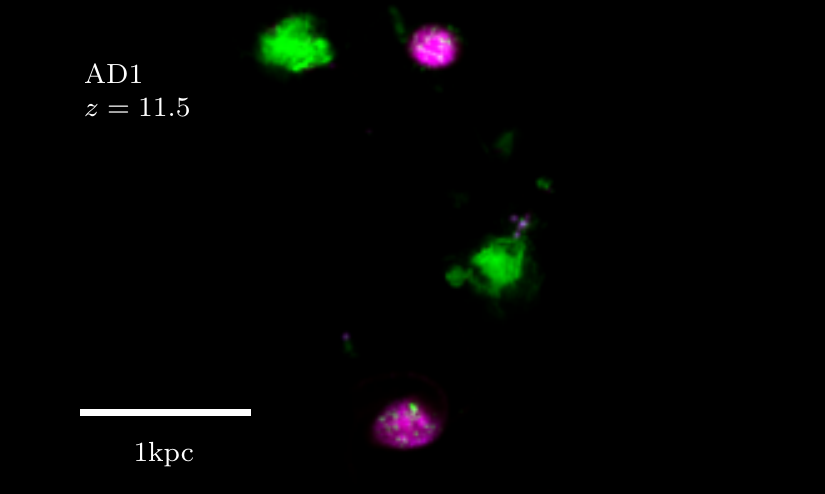}}
\centerline{\includegraphics[scale=1]{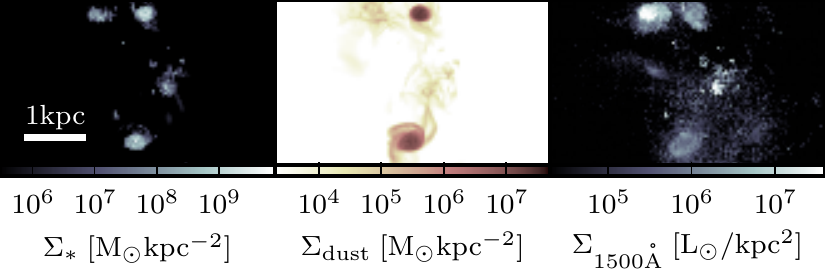}}
\caption{Demonstration of kpc-scale spatial offsets between [CII]~157.6$\mu$m and [OIII]~88.33$\mu$m during a merger.  (Top) Maps of HI column density, density weighted temperature, and metal mass surface density for AD1 at $z=11.5$ in a face-on view.  These have been computed in a 5kpc$\times$3kpc box surrounding the galaxy.  (Centre) Three channel image of [CII]~157.6$\mu$m (red), [OIII]~88.33$\mu$m (green), and [OI]~63.17$\mu$m (blue) showing the same region as in the top row.  This image is shown in linear scale (and normalised to 70\% of the maximum flux in each line). There are clear spatial offsets between [CII]~157.6$\mu$m/[OI]~63.17$\mu$m and [OIII]~88.33$\mu$m.  Note how the [OIII]~88.33$\mu$m emission corresponds to areas with reduced neutral gas content and higher temperature while the opposite is true for the [CII]~157.6$\mu$m/[OI]~63.17$\mu$m emitting regions. (Bottom) Maps of young stellar mass surface density, dust mass surface density, and 1500$\angstrom$ surface brightness.  Young stars are defined as having ages $<10$Myr.  Dust attenuation causes significant dimming around the locations of most of the young stars at 1500$\angstrom$.}
\label{spatoff}
\end{figure}

Spatial and spectral offsets have now been observed in numerous systems between far-infrared lines and UV/optical tracers of high-redshift galaxies (see Figure~6 of \citealt{Carniani2017} and references therein).  One of the best examples of this is BDF-3299 which exhibits kpc-scale spatial offsets between [CII]~157.6$\mu$m, [OIII]~88.33$\mu$m, and the UV as well as spectra offsets $\gg100$km/s \citep{Carniani2017}.  In general, many high-redshift galaxies appear to have clumpy emission \citep{Carniani2018}. There are seemingly no strong trends with the SFR of the clumps and interestingly, Ly$\alpha$ has even been observed as being blue-shifted with respect to the far-infrared lines (this is e.g. the case for JD1 \citealt{Hashimoto2018}).  However, spatial and spectral offsets are not ubiquitous. There are many examples where no offsets are found \citep[e.g.][]{Pentericci2016}.  Furthermore, these offsets occur not only in emission lines. Spatial offsets have also been observed between the dust continuum and the UV \citep{Bowler2018}.  However, a consensus has yet to be reached regarding the physical origins of these offsets.

We have searched our simulation outputs to determine whether spatial offsets occur between [CII]~157.6$\mu$m and [OIII]~88.33$\mu$m emission.  In Figure~\ref{spatoff}, we show an example of AD1 at $z=11.5$ of an RGB image where [CII]~157.6$\mu$m, [OIII]~88.33$\mu$m, and [OI]~63.17$\mu$m emission have been plotted in red, green, and blue, respectively, in linear scale.  Spatial offsets on kpc scales exist between the four clumps in the image, two of which are bright in [CII]~157.6$\mu$m/[OI]~63.17$\mu$m, and two of which are bright in [OIII]~88.33$\mu$m.  Among the simulation snapshots that we examined, spatial offsets between these IR lines are not ubiquitous and they both vary with time and viewing angle.  In order to generate these offsets, certain physical conditions are required: Firstly, the galaxy must be clumpy either due to a merger, the presence of small satellites, or a fragmented disk.  Secondly, to produce [OIII]~88.33$\mu$m without [CII]~157.6$\mu$m or [OI]~63.17$\mu$m, the neutral gas must be almost completely destroyed via photoionisation or SN feedback (see the top left panel of Figure~\ref{spatoff}).  However, the SN feedback cannot be so strong that the ionised intermediate density gas at $1\lesssim\rho/{\rm cm^{-3}}\lesssim 100$ with a temperature of $T\sim10^4-10^5$K has been completely disrupted.  Such ideal conditions are exhibited by AD1 at this particular redshift, but this phase is transient (see Figure~\ref{time_series}).  These offsets are no longer present at $z=11$.  

\begin{figure}
\centerline{\includegraphics[scale=1,trim={0 0.8cm 0 1cm},clip]{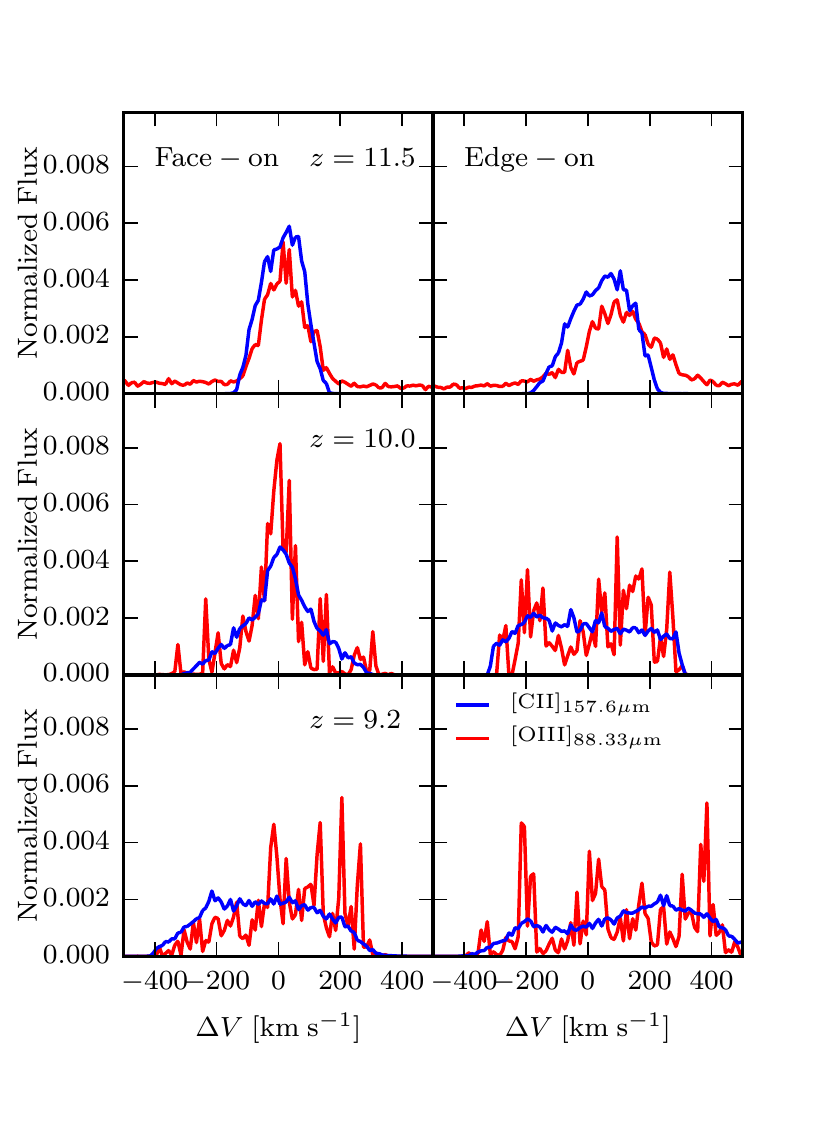}}
\caption{Normalised [CII]~157.6$\mu$m and [OIII]~88.33$\mu$m spectra for AD1 at three different redshifts (top row: $z=11.5$, middle row: $z=10.0$, bottom row: $z=9.2$).  The left and right columns show the face-on and edge-on views, respectively.}
\label{specoff}
\end{figure}

Since the [OIII]~88.33$\mu$m line is powered much closer to intense star-forming regions compared to [CII]~157.6$\mu$m, each line naturally adopts the velocity profiles of these different regions of phase-space.  In the top row of Figure~\ref{specoff}, we plot normalised [CII]~157.6$\mu$m and [OIII]~88.33$\mu$m spectra for AD1 at the same redshift in both face-on and edge-on configurations.  In the face-on view, the peak of the [CII]~157.6$\mu$m and [OIII]~88.33$\mu$m spectra seem to occur at similar locations.  When viewed edge-on, a small spectral offset appears between the two lines.  We also show the spectra at $z=10$ and $z=9.2$ and the spectral shapes are completely different. [OIII]~88.33$\mu$m emission tends to have a narrower profile compared to [CII]~157.6$\mu$m and it is often the case that the [OIII]~88.33$\mu$m spectra has multiple peaks, consistent with the velocities of individual star forming regions.  At $z=9.2$, AD1 is undergoing a merger which results in an even broader [CII]~157.6$\mu$m spectrum and potentially huge, $\gg100{\rm km\ s^{-1}}$ spectral offsets between [CII]~157.6$\mu$m and [OIII]~88.33$\mu$m.  The largest spectral offsets are a result of mergers and the presence of satellite galaxies, while smaller spectral offsets are due to [CII]~157.6$\mu$m and [OIII]~88.33$\mu$m probing different regions of temperature-density phase space that exhibit slightly different velocities.  In summary, neither spatial nor spectral offsets between [CII]~157.6$\mu$m and [OIII]~88.33$\mu$m are ubiquitous in our simulations but they can be easily reproduced at transient intervals.

\subsubsection{IR-SFR Relations}
\label{IR_SFR}
Since [CII]~157.6$\mu$m is expected to be one of the brightest far-infrared lines at high-redshift, it has been the target of many ALMA observations at $z>6$.  In many of these observations, [CII]~157.6$\mu$m has not been detected which raises the question of whether there are fundamental differences between the ISM of high-redshift galaxies compared to those in the local Universe.  Understanding why [CII]~157.6$\mu$m has not been observed for many galaxies is key to probing the ISM properties of observed systems.

\begin{figure}
\centerline{\includegraphics[scale=1]{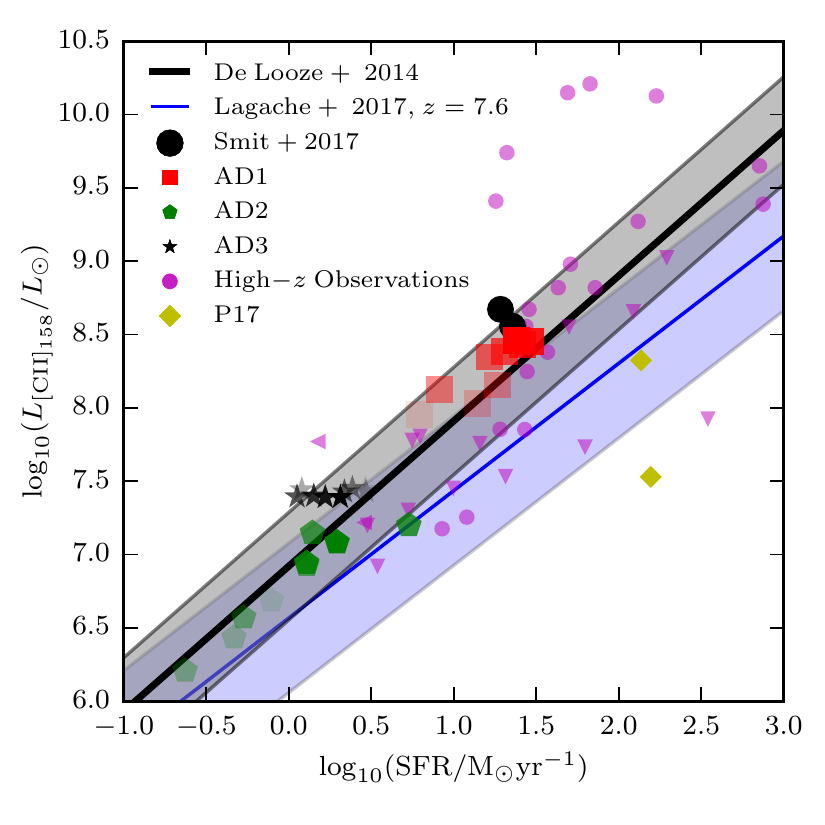}}
\caption{SFR versus [CII]~157.6$\mu$m luminosity for our three simulated galaxies as a function of redshift.  The squares, pentagons, and stars represent AD1, AD2, and AD3, respectively.  These points are coloured from light to dark depending on redshift with the lightest points being at $z=12$ and the darkest points being at $z=9.2$.  For comparison, we show the [CII]-SFR relation measured from $z=0$ galaxies from \protect\cite{DeLooze2014}, as well as numerous high-redshift galaxies and clumps compiled by \protect\cite{Carniani2018}.  Circles and triangles represent high-redshift observations.  Leftward and downward triangles indicate upper limits on the SFR and the $L_{\rm [CII]}$, respectively.  Yellow diamonds indicate the results from the hydrodynamics zoom-in simulations of \protect\cite{P17a,P17b}.  The black circles represent the observed systems from \protect\cite{Smit2018} that may be rotating.  The blue region shows the predicted relation and scatter from the simulations of \protect\cite{Lagache2018}.}
\label{CIISFR}
\end{figure}

In Figure~\ref{CIISFR}, we show the evolution of AD1, AD2, and AD3 in the [CII]-SFR plane between $12.0\geq z\geq9.2$ compared to the local relation \citep{DeLooze2014} as well as numerous observed systems collected in \cite{Carniani2018}.  Nearly all of our simulated points fall directly on the local relation.  There are many explanations for why deficits might occur:  Since [CII]~157.6$\mu$m predominantly emits from neutral gas, systems that are undergoing an intense burst of star formation would naturally be deficient of this gas phase.  However an intense burst of star formation would also make them the brightest in the UV and thus more easily observable. Indeed \cite{SilverRush2018} find a strong anti-correlation between [CII]~157.6$\mu$m luminosity and Ly$\alpha$ equivalent width.  One can see that for AD2, there is a point in the [CII]-SFR plane where the galaxy has undergone a burst of star formation which moves it to the right in the diagram due to the increase in SFR and also down on the diagram due to the reduction in neutral gas.  AD2 has the burstiest star formation history of the three galaxies so while this may make the galaxy more observable at specific times, we would naturally expect a deficit of [CII]~157.6$\mu$m emission during these bursts.  Thus we suggest that the [CII]~157.6$\mu$m deficit may be a selection bias due to selecting in the UV.

Two systems that were not selected in the UV are from \cite{Smit2018} (these were selected based on [OIII]5007$\angstrom$,4959$\angstrom+{\rm H\beta}$) and they are shown as the black circles in Figure~\ref{CIISFR}.  These two points fall much closer to the local relation which is more consistent with our predictions.   

Comparing AD2 with AD3, these systems have very similar halo mass, stellar mass, and metallicity at $z\sim9$, but different morphologies and star formation histories.  AD3 tends to fall higher on the relation than AD2 due to the different gas density distributions and chemical states of their ISMs.  We predict that the observed galaxies that fall closer to the $z=0$ relation from \cite{DeLooze2014} may be the more settled objects at high redshift.  This interpretation also agrees with that presented in \cite{Smit2018}.  Finally, we should also highlight the fact that some of the measurements which previously had upper limits on [CII]~157.6$\mu$m emission have been revised with newer observations that place them directly on the local [CII]-SFR relation \citep{Carniani2018b}, also relieving tension with our predictions.

\begin{figure}
\centerline{\includegraphics[scale=1]{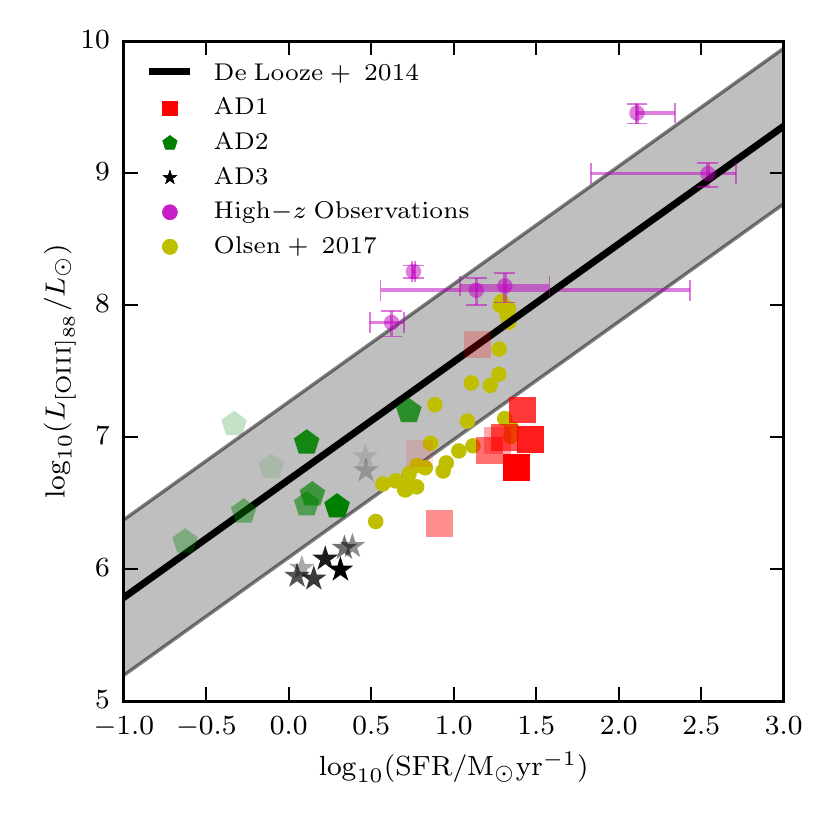}}
\caption{SFR versus [OIII]~88.33$\mu$m luminosity for our three simulated galaxies as a function of redshift.  The squares, pentagons, and stars represent AD1, AD2, and AD3, respectively.  These points are coloured from light to dark depending on redshift with the lightest points being at $z=12$ and the darkest points being at $z=9.2$.  For comparison, we show the [OIII]-SFR relation measured from $z=0$ galaxies from \protect\cite{DeLooze2014}, as well as numerous high-redshift galaxies from \protect\cite{Inoue2016,Laporte2017,Carniani2017,Tamura2018,Hashimoto2018,Hashimoto2018b}.  Yellow circles indicate the results from the hydrodynamics  simulations of \protect\cite{Olsen2017}.  }
\label{OIIISFR}
\end{figure}

Shown as yellow diamonds on Figure~\ref{CIISFR} are the results from hydrodynamic zoom-in simulations from \cite{P17a,P17b} which have a very similar halo mass as AD1.  Both of these simulated systems show a large deficit in [CII]~157.6$\mu$m emission compared to the local relation.  Our simulations use very different models for stellar feedback, compared to \cite{P17a,P17b}.  We also include on-the-fly radiative transfer and have higher resolution which may help explain the differences.  The stellar feedback in our simulation is tuned such that the stellar masses of the objects are consistent with the extrapolated predictions from abundance matching \citep{Behroozi2013} and if these relations are representative of high-redshift galaxies, the \cite{P17a,P17b} galaxies would have too high a stellar mass and SFR to be consistent with abundance matching.  This artificially shifts the galaxies to the right on the diagram leading to a [CII]~157.6$\mu$m deficit and provides a natural explanation for the disagreement between our study and theirs.

Finally, we also show the fitted $z=7.6$ [CII]-SFR relation from the simulations of \cite{Lagache2018} as the blue band.  This also has a [CII]~157.6$\mu$m deficit compared to the local relation, albeit with large scatter.  The authors indicate that this deficit is strongly correlated with the intensity of the local radiation field so modelling this correctly is certainly a crucial ingredient for making accurate predictions.  The presence of self-shielded regions with low levels of flux may help boost the [CII]~157.6$\mu$m luminosity.  Such regions are often present in AD1 and AD3, but not always AD2 which may indicate why AD2 is in agreement with the \cite{Lagache2018} model but AD1 and AD3 fall above.  Note however that the radiation fields were modelled very differently in \cite{Lagache2018} compared to our work.

[OIII]~88.33$\mu$m emission has now been observed in a few high-redshift galaxies and thus we can compare our simulations with both the local relation with SFR from \cite{DeLooze2014} and the [OIII]~88.33$\mu$m luminosities of high-redshift galaxies (see Figure~\ref{OIIISFR}).  In general, we find that our predicted [OIII]~88.33$\mu$m luminosities tend to fall below both the local relation and that observed at high redshifts.  A similar trend to ours was found in the simulations of \cite{Olsen2017}, yet the simulations of \cite{Moriwaki2018} tend to find the opposite in that their galaxies fall above the local relation.  \cite{Katz2018} discussed a few reasons why our simulated systems may fall below the observed relation and pointed out that we use Solar abundance ratios \citep{Grevesse2010} of [O/Fe] when computing the [OIII]~88.33$\mu$m luminosity with {\small CLOUDY}; however, high-redshift galaxies are likely predominantly enriched by type-II core-collapse supernova which would lead to an enhancement of oxygen at fixed metallicity compared to Solar.  This was also proposed by \cite{Steidel2016} to help reconcile spectral modelling of $z\sim2-3$ galaxies.  Observations of [OIII]~88.33$\mu$m are still scant at high redshift so more data will be needed to conclude if there is indeed a disagreement between our models and high-redshift galaxies in this property.  Furthermore, a consensus has yet to be reached between different simulations of what [OIII]~88.33$\mu$m luminosities should be expected for a given SFR at high-redshift.

\begin{figure}
\centerline{\includegraphics[scale=1]{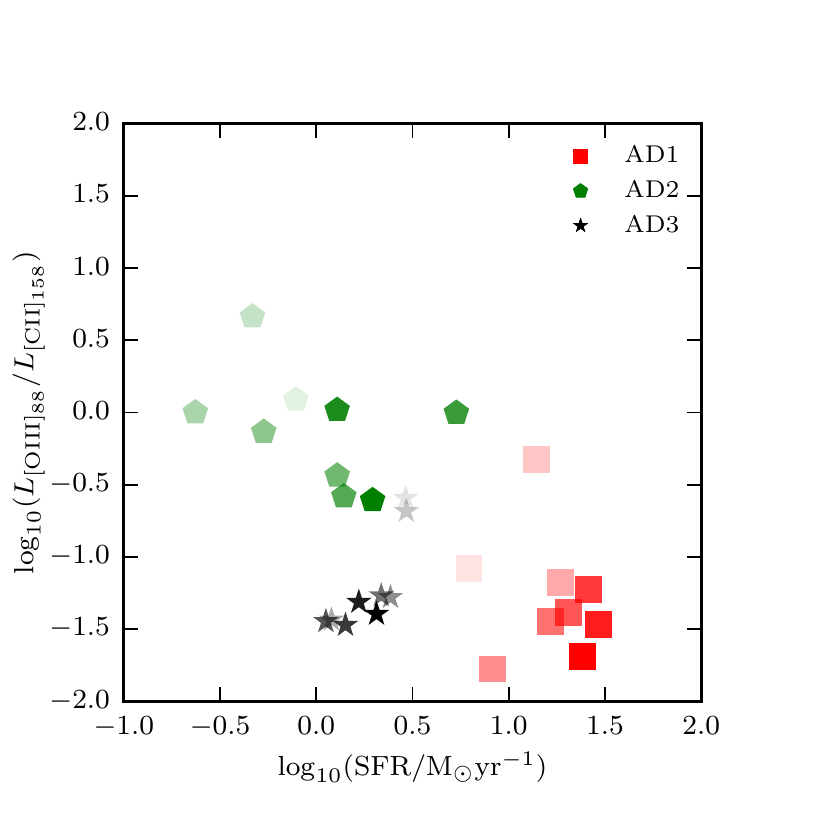}}
\caption{SFR versus [OIII]~88.33$\mu$m/[CII]~157.6$\mu$m luminosity for our three simulated galaxies as a function of redshift.  The squares, pentagons, and stars represent AD1, AD2, and AD3, respectively.  These points are coloured from light to dark depending on redshift with the lightest points being at $z=12$ and the darkest points being at $z=9.2$.}
\label{CIIOIII}
\end{figure}

\cite{Hashimoto2018b,Hashimoto2018c} have pointed out that there exists a negative correlation between the ratio of [OIII]~88.33$\mu$m to [CII]~157.6$\mu$m luminosity and the bolometric luminosity of the host.  The bolometric luminosity is expected to scale with the SFR of the galaxy.  In Figure~\ref{CIIOIII}, we plot [OIII]~88.33$\mu$m/[CII]~157.6$\mu$m versus SFR for AD1, AD2, and AD3, at multiple redshifts.  We also see a negative correlation between this ratio and the SFR for our systems.  The SFR also correlates with halo mass and it may be the case that more massive haloes are less susceptible to SN feedback which drives [OIII]~88.33$\mu$m emission and destroys the neutral gas where [CII]~157.6$\mu$m originates.  Hence the galaxies that are less sensitive to SN feedback are expected to be more massive (and have higher bolometric luminosity) which would decrease [OIII]~88.33$\mu$m/[CII]~157.6$\mu$m.  This argument breaks down when a bright AGN is present as the hard radiation field can increase the amount of [OIII]~88.33$\mu$m. Indeed, \cite{Hashimoto2018c} finds that one of their two high-redshift quasars has a [OIII]~88.33$\mu$m/[CII]~157.6$\mu$m ratio that is more consistent with galaxies that have a bolometric luminosity that is lower by two orders of magnitude.

\subsection{Nebular Lines}
Moving from the IR to shorter wavelengths, UV and optical nebular lines can provide an additional and complementary probe of the ISM of high-redshift galaxies.  While our current understanding of high-redshift nebular lines comes from their presence in broadband filters \citep[e.g.][]{Stark2013}, the impending launch of JWST will provide much more detailed information.  For this reason, predicting their properties is a highly relevant exercise.  Furthermore, numerous unexplained properties of $z\sim2-3$ galaxy nebular lines persist so it is useful to see how the much higher-redshift systems compare to these intermediate-redshift galaxies.

\begin{figure*}
\centerline{\includegraphics[scale=1]{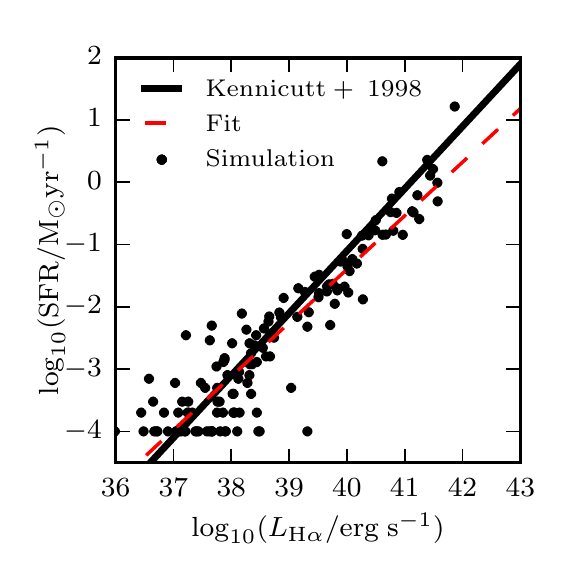}\includegraphics[scale=1]{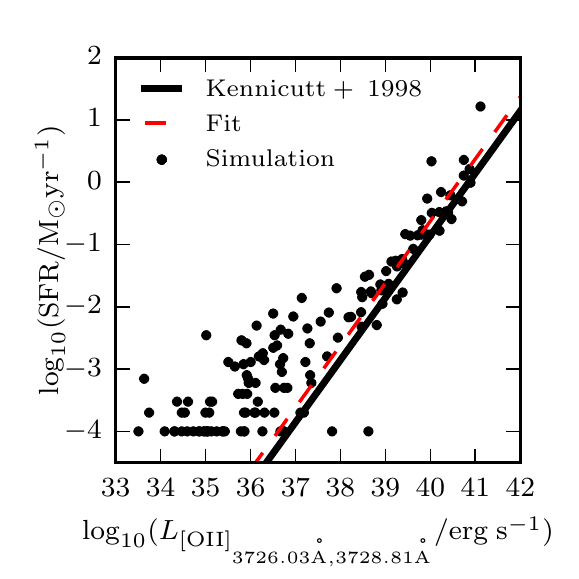}\includegraphics[scale=1]{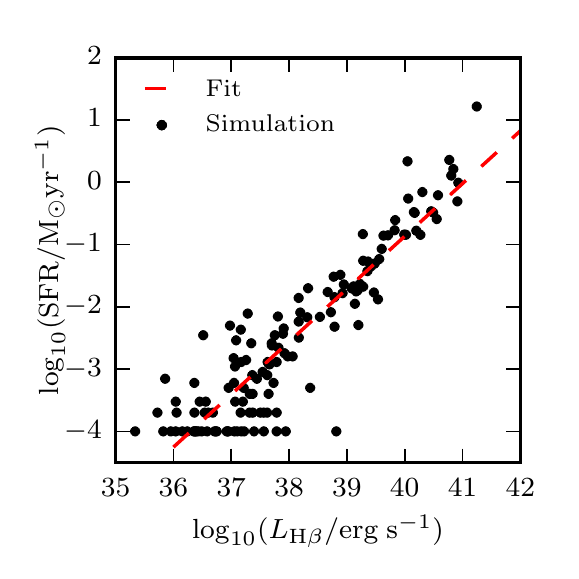}}
\caption{SFR versus H$\alpha$ (Left), [OII]~3727$\angstrom$ doublet (Centre), and H$\beta$ (Right) are shown for all of our simulated galaxies at $z=10$ that contain star formation.  The solid black lines show the fits from \protect\cite{Kennicutt1998a}, while the dashed red line shows the fit from our simulation.  Since the empirical [OII]-SFR relation is not a strict power-law, our fit only applies to $L_{\rm [OII]}>10^{38}\ {\rm erg\ s^{-1}}$.  The slopes of our fitted relations are consistent with \protect\cite{Kennicutt1998a} and the offsets are likely due to effects of metallicity, different intrinsic SEDs, and LyC leakage.  The lines of points at low SFR is a result of the stellar mass resolution of our simulation.}
\label{neb_sfr}
\end{figure*}

\subsubsection{Nebular Lines as SFR Indicators at High Redshift}
Recombination lines, in particular H$\alpha$, are good indicators of the SFR of a galaxy because they are directly sensitive to the number of ionising photons which are primarily emitted by young massive stars.  For an assumed metallicity and stellar IMF, one can calculate the total number of emitted ionising photons for a simple stellar population of a given mass and compare this with the number of ionisations needed to drive a specific H$\alpha$ luminosity for a given recombination rate (usually Case~B) and gas temperature (often 10,000K).  In the left panel of Figure~\ref{neb_sfr}, we show the intrinsic H$\alpha$ luminosity for all star-forming galaxies at $z=10$ compared with the SFR averaged over the previous 10Myr.  As expected, the H$\alpha$ luminosity shows a reasonably tight power-law relation with the total SFR of the galaxy.  This provides an additional consistency check that our newly developed machine learning method for extracting the nebular line luminosities on a cell-by-cell basis within the simulation is performing well.  The solid black line shows the H$\alpha$-SFR calibration from \cite{Kennicutt1998a} and this indeed provides a good first approximation of the SFRs of our simulated systems.  At $L_{\rm H\alpha}\gtrsim10^{39}\ {\rm erg\ s^{-1}}$, the majority of our simulated galaxies tend to scatter to higher H$\alpha$ luminosities for a given SFR.  The \cite{Kennicutt1998a} calibration is based on a Salpeter IMF at solar metallicity while in our simulation, we use the BPASS300 model which produces more ionising photons for the same mass stellar cluster.  Furthermore, the stars in our galaxy are at lower metallicity which would also help produce more ionising photons compared to what was used in \cite{Kennicutt1998a}.  Both of these effects naturally drive our systems to have higher H$\alpha$ luminosities for a fixed SFR.  However, it is generally assumed when calibrating the H$\alpha$-SFR relation that all ionising photons produce an ionisation whereas in our simulations, the LyC escape fraction is not necessarily zero for all of our systems and the presence of H$_2$, He, and dust may also absorb ionising photons.  The combination of these effects introduce scatter into the relation while also decreasing the H$\alpha$ luminosity.  We have fit our simulated galaxies to calibrate the high-redshift H$\alpha$-SFR relation and find:
\begin{equation}
\log_{10}({\rm SFR/M_{\odot}\ yr^{-1}})=0.86\log_{10}(L_{\rm H\alpha}/{\rm erg\ s^{-1}})-35.81,
\end{equation}
which is slightly shallower than the \cite{Kennicutt1998a} calibration.

A similar exercise can be done using the H$\beta$ recombination line as shown in the right panel of Figure~\ref{neb_sfr}.  This line is not generally used as much in the local Universe compared to H$\alpha$ because it is weaker and more susceptible to stellar absorption.  However, JWST will be able to observe both H$\alpha$ and H$\beta$ at high redshift so having a calibrated relation is useful.  In our models, we also find a power-law relation such that:
\begin{equation}
\log_{10}({\rm SFR/M_{\odot}\ yr^{-1}})=0.85\log_{10}(L_{\rm H\beta}/{\rm erg\ s^{-1}})-34.73.
\end{equation}

This relation is remarkably similar to the H$\alpha$ calibration in terms of slope but has a normalisation that gives an order of magnitude weaker luminosity at fixed SFR compared to H$\alpha$.

In the lower redshift Universe, forbidden lines have also been used as SFR indicators, including the [OII]~3726.03$\angstrom$, 3728.81$\angstrom$ doublet.  However, these lines are not directly sensitive to the overall ionising luminosity of the host stellar population and thus need to be calibrated empirically.  In the middle panel of Figure~\ref{neb_sfr}, we plot the [OII] luminosity against the SFR for our simulated $z=10$ galaxies and compare with the calibration from \cite{Kennicutt1998a}.  While at higher luminosities ($L_{\rm [OII]}\gtrsim10^{38}\ {\rm erg\ s^{-1}}$), the slope is very similar to the local relation, we see an offset such that at fixed SFR, we have lower $L_{\rm [OII]}$.  This is very likely due to the fact that our simulated high-redshift galaxies have much lower metallicities compared to low-redshift galaxies.  Furthermore, our simulated systems do not exhibit a strict power-law behaviour in this plane.  At lower luminosities ($L_{\rm [OII]}\lesssim10^{38}\ {\rm erg\ s^{-1}}$), the galaxies tend to exhibit even weaker $L_{\rm [OII]}$ at fixed SFR compared to \cite{Kennicutt1998a}.  Since many of these systems have had very little star formation, this may be a result of a steepening SFR-metallicity relation. Since JWST will likely only be able to image the more luminous systems, we can calibrate the high-redshift $L_{\rm [OII]}$-SFR relation by only fitting those systems with $L_{\rm [OII]}>10^{38}\ {\rm erg\ s^{-1}}$.  We find:
\begin{equation}
\log_{10}({\rm SFR/M_{\odot}\ yr^{-1}})=1.00\log_{10}(L_{\rm [OII]}/{\rm erg\ s^{-1}})-40.55.
\end{equation}
The slope of this relation is unity, consistent with \cite{Kennicutt1998a}, but has a normalisation that is slightly lower.

\subsubsection{High-Redshift Diagnostic Diagrams}
Line ratios and diagnostic diagrams are used to disentangle properties of galaxies such as the sources of the radiation \citep[e.g.][]{Baldwin1981} or the metallicity \citep[e.g.][]{Pagel1979,McGaugh1991}.  For reasons discussed earlier, being able to disentangle star-forming galaxies from AGN at high redshifts is particularly important, especially for measuring the contribution of AGN to reionization.   Historically, the BPT diagram has been the primary discriminator for low redshift galaxies \citep{Baldwin1981,Kewley2001,Kauffmann2003}.  However, intermediate redshift galaxies ($z\sim2-3$) have shown clear offsets in this diagram compared with the low-redshift locus \citep[e.g.][]{Strom2017} and thus it is crucial to understand both the reason for this offset and what behaviour high-redshift galaxies are expected to have in this plane.

\begin{figure*}
\centerline{\includegraphics[scale=1]{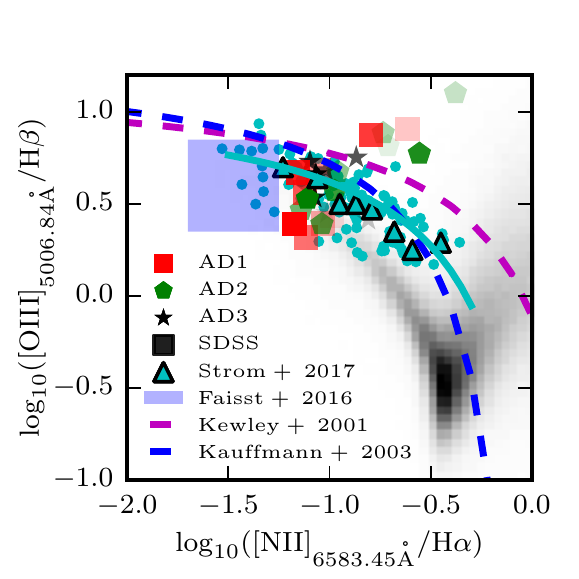}\includegraphics[scale=1]{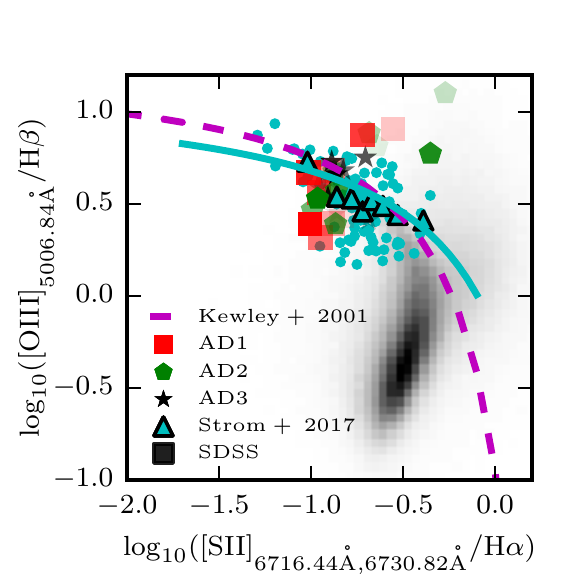}\includegraphics[scale=1]{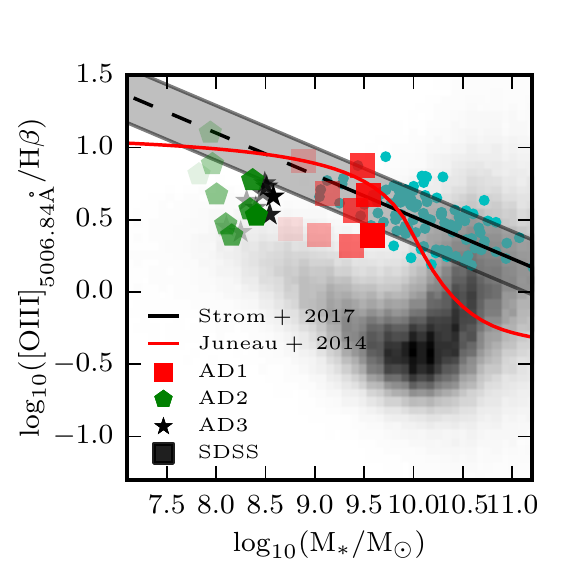}}
\centerline{\includegraphics[scale=1]{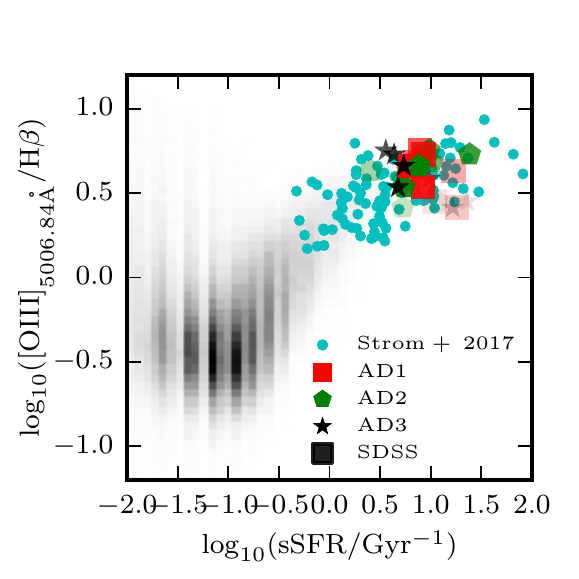}\includegraphics[scale=1]{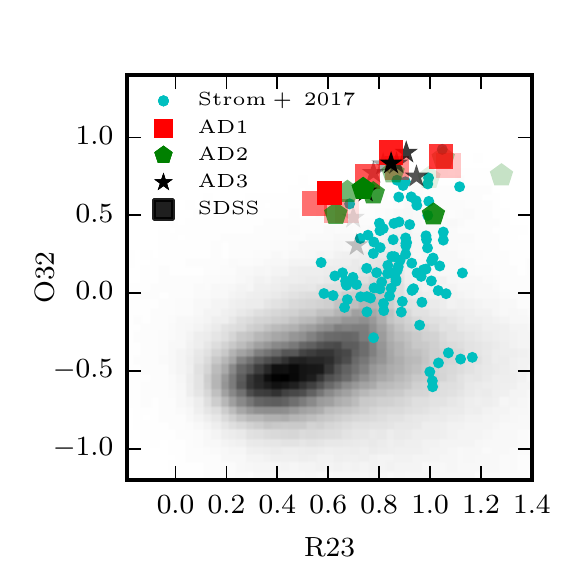}\includegraphics[scale=1]{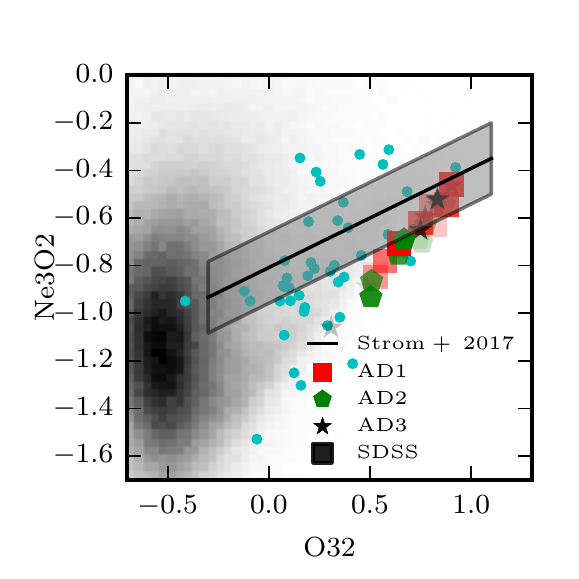}}
\caption{Strong line diagnostic diagrams.  (Top Left) [NII]-BPT diagram.  Our three massive systems AD1, AD2, and AD3 are shown as the red square, green pentagon, and black star, respectively, with more translucent points representing higher redshifts. The purple and blue dashed lines show the AGN separation lines from \protect\cite{Kewley2001} and \protect\cite{Kauffmann2003}, respectively.  The 2D grey histogram shows the local SDSS galaxies \protect\citep{Thomas2013} while the cyan triangles and lines show the binned median line ratios and the fit to these medians from \protect\cite{Strom2017}.  Small cyan circles represent individual galaxies from KBSS \protect\citep{Strom2017}.  The large blue square shows the location of $z=6$ galaxies from \protect\cite{Faisst2016} assuming the [NII]/H$\alpha$ versus metallicity relation from \protect\cite{Maiolino2008}. (Top Centre) [SII]-BPT diagram.  All symbols and lines are the same as in the top left panel.  (Top Right)  The mass-excitation relation showing the stellar mass of the galaxy versus [OIII]~5007$\angstrom/{\rm H\beta}$.  SDSS galaxies are shown in greyscale.  The red line demarcates the boundary between star-forming galaxies and AGN in the local Universe from \protect\cite{Juneau2014}.  The black line and surrounding grey region shows the fit to the KBSS galaxies and the 1$\sigma$ standard deviation around the fit.  The dashed part of this line shows where the fit has been extrapolated.  (Bottom Left) sSFR-excitation relation.  Symbols are the same as in the top right panel.  (Bottom Centre) R23 versus O32 for SDSS galaxies shown in greyscale compared to our simulated galaxies.  (Bottom Right) O32 versus Ne3O2 for KBSS galaxies compared with our simulated systems and SDSS. The black line and surrounding grey region shows the fit to the KBSS galaxies and the 1$\sigma$ standard deviation around the fit. }
\label{N2S2_BPT}
\end{figure*}

In the top left and top centre panels of Figure~\ref{N2S2_BPT}, we compare the location of AD1, AD2, and AD3 on the [NII] and [SII] BPT diagrams at different redshifts with that of low-redshift SDSS galaxies \citep{Thomas2013}, shown as the 2D grey histograms, as well as those at $z\sim2-3$ from the Keck Baryonic Structure Survey \citep[KBSS,][]{Strom2017}.  In general, our systems fall below the dividing line between star forming galaxies and AGN presented in  \cite{Kewley2001} and \cite{Kauffmann2003} although a few simulated points scatter above.  We have checked these specific systems and find a strong enhancement in star formation.  The increase in gas phase metallicity and the temperature of the gas, along with the burst of ionising radiation pushes the systems above the line.  

Our galaxies lie at the tail end of the SDSS distribution and although there is significant scatter in the [NII]-BPT diagram, there are many examples where our galaxies fall above the SDSS relation.  Enhancements in [OIII]~5007$\angstrom/{\rm H\beta}$ compared to SDSS have already been confirmed out to $z=5.5$ using COSMOS data \citep{Faisst2016}.  Assuming that these high-redshift COSMOS galaxies follow the [NII]/H$\alpha$ versus metallicity relation from \cite{Maiolino2008}, \cite{Faisst2016} make predictions for the expected location of $z\sim6$ galaxies on the BPT diagram.  These predictions are shown as the blue square on the top left panel of Figure~\ref{N2S2_BPT}.  While our simulated galaxies show a systematic offset compared to this prediction towards higher [NII]~6583$\angstrom$/H$\alpha$, the predictions for [OIII]~5007$\angstrom/{\rm H\beta}$ are in reasonable agreement.

\cite{Barrow2017} have also made predictions for the locations of high-redshift galaxies on the [NII] and [SII] BPT diagrams using the Renaissance simulations.  These simulations include all of the necessary radiative transfer, stellar feedback, and metal enrichment, similar to our simulations. However, while \cite{Barrow2017} use slightly higher resolution, they study lower mass objects.  Offsets in the BPT diagrams are also found in their simulations; however, many of the galaxies tend to scatter far below the SDSS relation rather than above.  Although they are not shown, we have checked the lower mass objects in our simulations and found that they also scatter to a similar area of the BPT diagram as \cite{Barrow2017} and this effect is likely due to very low metallicity for the lowest mass systems.

\begin{figure*}
\centerline{\includegraphics[scale=1]{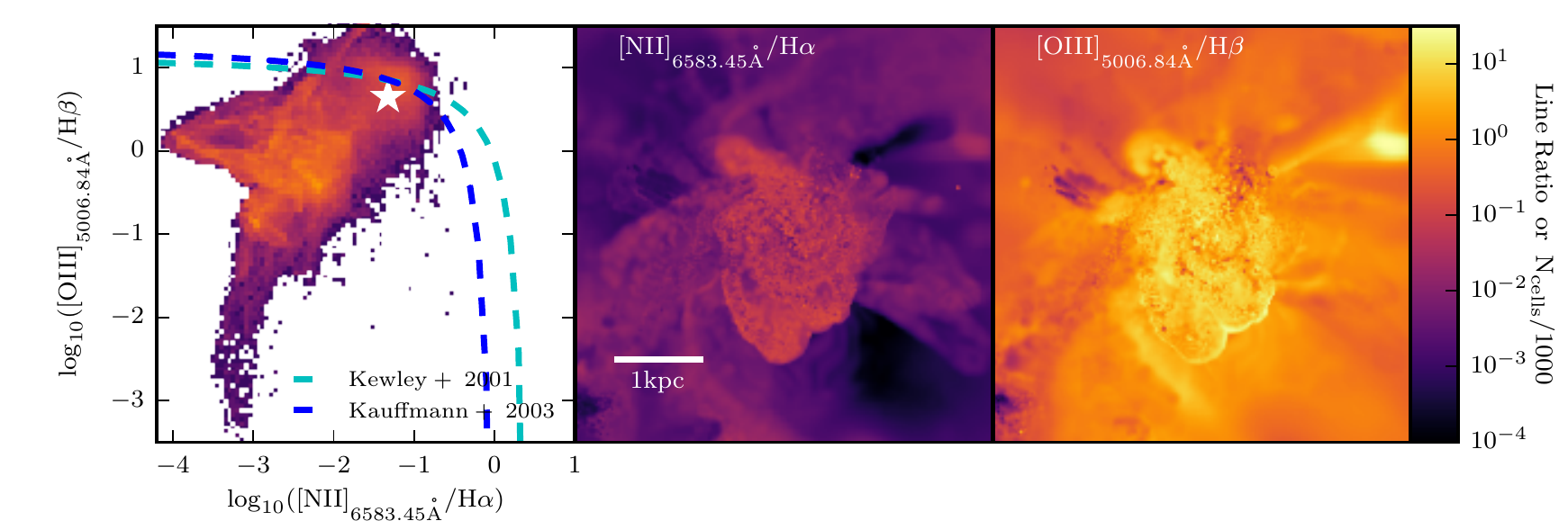}}
\caption{(Left) Spatially resolved [NII]-BPT diagram for the central 5kpc of AD1 at $z=10$.  The 2D histogram shows the location of all cells within this region while the white star represents the luminosity-weighted average of the galaxy.  The cyan and blue dashed lines show the AGN separation lines from \protect\cite{Kewley2001} and \protect\cite{Kauffmann2003}, respectively.  (Centre) Spatial map of [NII]~6583$\angstrom$/H$\alpha$ for the face-on view of AD1 at $z=10$.  (Right) Same as the centre panel but for [OIII]~5007$\angstrom/{\rm H\beta}$.}
\label{BPT3P}
\end{figure*}

\cite{Strom2017} have argued that the $z\sim2-3$ BPT offsets are likely due to higher sSFRs and harder ionising spectra from iron-poor stellar populations that may include massive binaries.  Our systems have SFRs at the lower end of their sample, similar sSFRs, and lower metallicity, but crucially, we employ the BPASS300 model for the stellar SED.  Changing the nebular metallicity only seems to result in a horizontal shift in the BPT diagram at fixed ionisation parameter (see Figure~26 of \citealt{Strom2017}). The other properties of our simulated systems are similar to the KBSS sample and our simulations may therefore provide a test of this hypothesis.  It is clear from the top left panel of Figure~\ref{N2S2_BPT} that AD1, AD2, and AD3 scatter around the median points of the KBSS sample, which would agree with the hypothesis that harder ionising spectra from iron-poor stellar populations that include massive binaries can result in the BPT offset.  

The question that still remains is how similar our simulated systems are to the KBSS galaxies considering ours are at significantly higher redshift.  There are multiple mechanisms for generating the BPT offsets including higher electron densities, enhanced N/O at fixed O/H, AGN activity, or harder spectra from metal poor stars that may include massive binaries.  We can compare the similarity between our systems and the KBSS galaxies by studying further diagnostic diagrams.  In top right and bottom left panels of Figure~\ref{N2S2_BPT}, we show the mass-excitation relation (M$_*$ versus [OIII]~5007$\angstrom/{\rm H\beta}$) and the sSFR-excitation relation (sSFR versus [OIII]~5007$\angstrom/{\rm H\beta}$) for AD1, AD2, and AD3 compared with local SDSS galaxies and the expectations from the KBSS sample.  In both diagnostics, our simulated systems show significant offsets from the SDSS sample, consistent with the KBSS galaxies.  The mass-excitation relation has the advantage of only requiring two emission lines, [OIII]~5007$\angstrom$ and ${\rm H\beta}$, compared with the BPT diagram which requires four.  \cite{Juneau2014} have demonstrated that this diagram can be used to select AGN versus star-forming galaxies and this is shown as the red line in the top right panel of Figure~\ref{N2S2_BPT}.  Similar to the BPT diagram, most of our simulated galaxies fall below this line and thus would be correctly classified using this discriminator.  In general, the [OIII]~5007$\angstrom$ and ${\rm H\beta}$ lines are brighter than [NII]~6583$\angstrom$ and are thus more easily observable at high redshift.  Hence, in the absence of [NII]~6583$\angstrom$ or H$\alpha$, the mass-excitation relation may provide an alternative diagnostic for classifying AGN at even the highest redshifts.

The bottom left panel of Figure~\ref{N2S2_BPT} shows the sSFR-excitation relation.  In contrast to the mass-excitation relation, \cite{Strom2017} find that the KBSS galaxies follow the trend of the local relation but exhibit higher sSFRs.  These results are consistent with slightly lower redshift galaxies from the 3D-HST survey \citep{Dickey2016} as well as $z\sim3.5$ Lyman-break-selected galaxies \citep{Holden2016}.  It is clear from this Figure that our galaxies also follow this trend, consistent with the observational data between $z\sim2-3.5$.

Continuing our analysis of nebular diagnostic diagrams, in the bottom centre panel of Figure~\ref{N2S2_BPT} we show R23 ($\log_{10}(([{\rm OIII}]\ 4960\angstrom,\ 5007\angstrom+[{\rm OII}]\ 3727\angstrom,\ 3729\angstrom)/{\rm H\beta})$) versus O32 ($\log_{10}({\rm [OIII]\ 4960\angstrom,\ 5007\angstrom/[OII]\ 3727\angstrom,\ 3729\angstrom})$), and in the bottom right panel, we show O32 versus Ne3O2 ($\log_{10}({\rm [NeIII]\ 3869\angstrom/[OII]\ 3727\angstrom,\ 3729\angstrom})$).  We again find good agreement between our simulated galaxies and the KBSS sample for both diagnostic diagrams.  The simulated galaxies are reasonably consistent with the trend of SDSS galaxies towards high R23, which was also seen in the $z\sim2.3$ galaxies from the MOSDEF survey \citep{Sanders2016}; although, the simulations have a slight tendency to fall to the left of this relation which can likely be ascribed to low metallicity (see Figure~24 of \citealt{Strom2017}).  Their location on this plane is indicative of higher excitation compared to SDSS systems.  Similarly, we also find that the simulated galaxies fall on the extended trend of the SDSS galaxies in the O32 versus Ne3O2 plane towards high values of O32 which is often used as an indicator of the ionisation parameter.

For all six diagnostic diagrams presented in this work, our simulated galaxies are very consistent with the intermediate redshift samples from KBSS and other surveys.  Because of this, we argue that similar diagnostics can be used at high redshifts to constrain the properties of the ISM.

\begin{figure*}
\centerline{\includegraphics[scale=1]{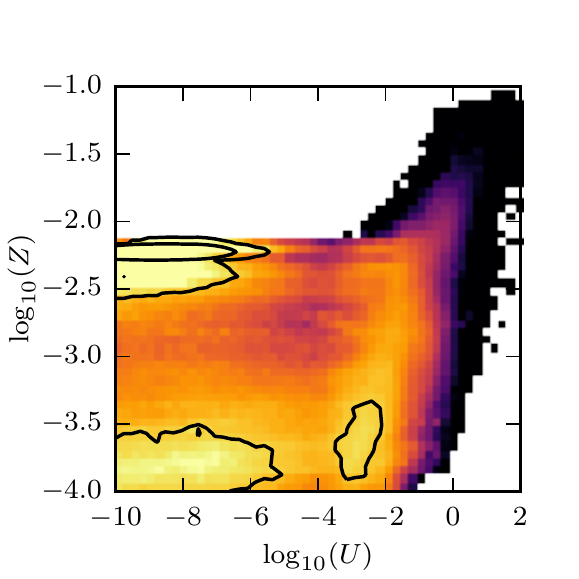}\includegraphics[scale=1]{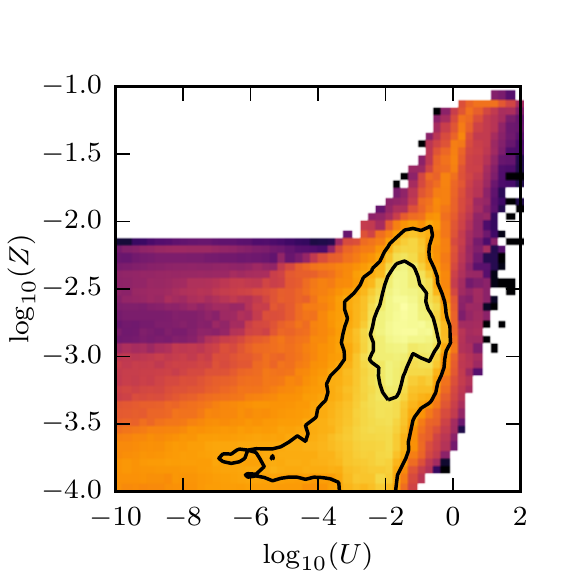}\includegraphics[scale=1]{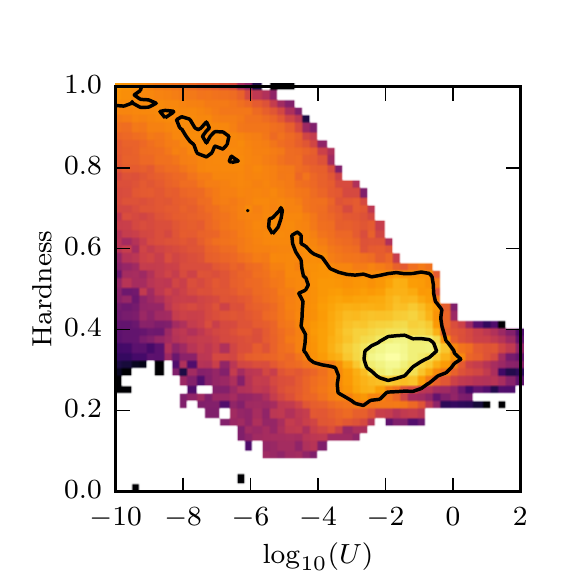}}
\caption{(Left) 2D mass-weighted histogram of ionisation parameter versus gas phase metallicity for all cells within the virial radius of AD1 at $z=10$.  (Centre) 2D volume-weighted histogram of ionisation parameter versus gas phase metallicity for all cells within the virial radius of AD1 at $z=10$. (Right) 2D volume-weighted histogram of ionisation parameter versus the hardness parameter for all cells within the virial radius of AD1 at $z=10$.  On all panels, contours indicate regions containing 50\% and 90\% of the total mass or volume.}
\label{iparam}
\end{figure*}

\subsubsection{Nebular Line Diversity Within Galaxies}
Thanks to large IFU surveys, it is now possible to separate line diagnostics for individual regions within the same galaxies for a large galaxy sample \citep[e.g.][]{Belfiore2016}.  Generally, when interpreting galaxy line diagnostics, assumptions must be made for properties such as the electron density, ionisation parameter, temperature, spectrum, and metallicity, despite the fact that these properties can vary by orders of magnitude throughout the individual galaxy.  In Figure~\ref{BPT3P}, we show the [NII]-BPT diagram for AD1 at $z=10$ where in the left panel, we plot a 2D histogram of [NII]~6583$\angstrom$/H$\alpha$ versus [OIII]~5007$\angstrom/{\rm H\beta}$ for all cells in the central 5kpc.  While the white star indicates the luminosity-weighted average of the entire galaxy, one can see that the individual conditions within a simulation cell allow the cells to scatter all around the BPT diagram.  Summarising this behaviour is obviously problematic.  

In the middle and right panel of Figure~\ref{BPT3P}, we show face-on maps of AD1 where the images are illuminated by either [NII]~6583$\angstrom$/H$\alpha$ (centre) or [OIII]~5007$\angstrom/{\rm H\beta}$ (right).  [NII]~6583$\angstrom$/H$\alpha$ is highest in the disk regions of the galaxy where the metallicity is also the highest.  This is also true for [OIII]~5007$\angstrom/{\rm H\beta}$; however, there are also prominent features in this image such as the shock in the lower right part of the disk.

One of the properties that can shift the location of a galaxy on the BPT and other diagnostic diagrams is the ionisation parameter, $U\equiv \frac{\Phi}{cn_{\rm H}}$.  $U$ is a dimensionless quantity where, $\Phi$ is the ionising photon flux in photons per unit area per unit time, $c$ is the speed of light, and $n_{\rm H}$ is the number density of hydrogen atoms.  To give an idea of the change in $U$ throughout the galaxy, for AD1 at $z=10$, we find that the mean volume-weighted ionisation parameter is $4.2\times10^{-2}$ while the mean mass-weighted ionisation parameter is $6.7\times10^{-4}$.  In the left panel of Figure~\ref{iparam}, we show a 2D mass-weighted histogram of ionisation parameter versus gas phase metallicity for all cells within the virial radius of AD1 at $z=10$.  Most of the gas mass has very low ionisation parameter but is metal enriched.  This is representative of the cold neutral gas disk at the centre of the halo.  In contrast, the middle panel shows the same 2D histogram, but in this case, it has been volume-weighted. Much of the volume of the galaxy exists at significantly higher ionisation parameter.  This is more representative of the low density halo gas around the central disk.  There is a very strong tail in the distribution towards very high ionisation parameter and metallicity which is representative of the regions around young stars that have just gone SN.  The ionisation parameter remains high because in the BPASS SED, ionising photons can be emitted long after the first SN in the cluster have gone off and the metal enrichment from the SN makes this gas have highly super-Solar metallicities.  As we have seen from Figure~\ref{BPT3P}, the diversity within the galaxy is huge and in this case, the ionisation parameter can change by many orders of magnitude so it would be very difficult to describe the galaxy with a single value.

In addition to the ionisation parameter, the hardness of the SED can influence the luminosities of nebular lines.  In particular, the harder the SED, the brighter we would expect the higher ionisation potential lines to be.  In the right panel of Figure~\ref{iparam}, we show a 2D volume-weighted histogram of ionisation parameter versus the hardness parameter which we define as $N_{\gamma}(E\geq24.59{\rm eV})/ N_{\gamma}(E\geq13.60{\rm eV})$, where $N_{\gamma}(E)$ is the number density of photons in a simulation cell with energy $>E$.  For our chosen BPASS SED at the mean metallicity of AD1 at $z=10$, the hardness parameter near star particles is expected to be $\sim0.3$, consistent with the right panel of Figure~\ref{iparam}.  However, we can once again see that the hardness of the spectrum changes throughout the galaxy.  At high values of the ionisation parameter, one is probing the regions near young stars and the hardness parameter drops to values between $0.2-0.4$, as one would expect for our chosen SED.  As the ionisation parameter decreases, the hardness of the photons in the cells increases because the lower energy photons are preferentially absorbed by the hydrogen\footnote{Note that this effect is only crudely captured by our simulations because of the coarse multi-frequency radiation bins.}.  In other words, the galaxy is more optically thin to photons with $E=24.59$eV compared to $E=13.6$eV.  Thus depending on where it is in the galaxy, a gas cell will have a different temperature, density, metallicity, ionisation parameter, and hardness parameter.  Furthermore, these parameters may change between the regions that are brighter in different lines.  These inhomogeneities are not unique to our high-redshift galaxies and should be present in galaxies down to the present epoch.

\begin{figure*}
\centerline{\includegraphics[scale=1]{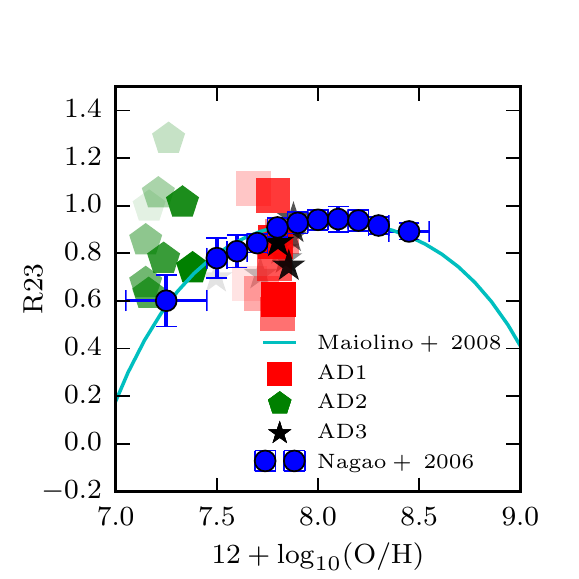}\includegraphics[scale=1]{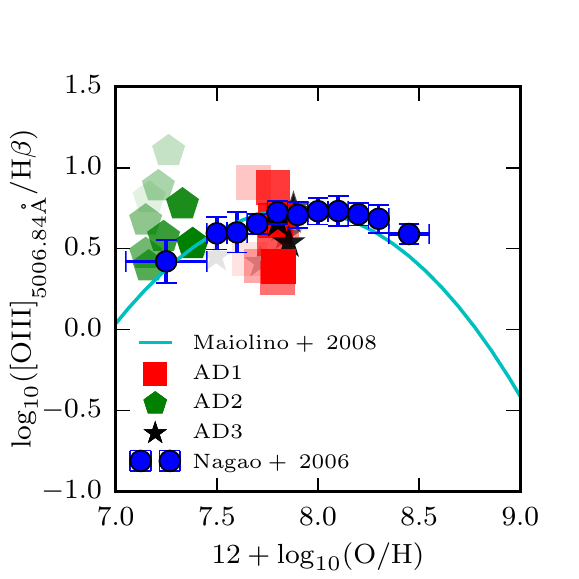}\includegraphics[scale=1]{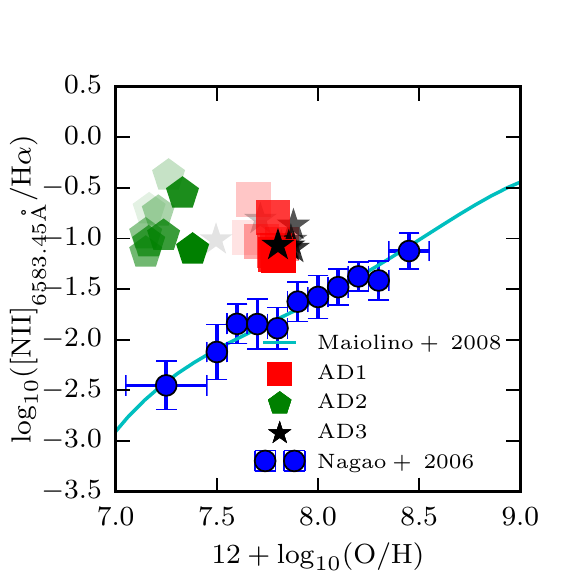}}
\caption{Nebular line metallicity indicators.  We show $12+\log_{10}({\rm O/H})$ versus R23 (Left), versus [OIII]~5007$\angstrom/{\rm H\beta}$ (Centre) and versus [NII]~6583$\angstrom$/H$\alpha$ (Right).  Red, green, and black points represent AD1, AD2, and AD3 , respectively.  Cyan lines are fitted relations from \protect\cite{Maiolino2008} which were derived from low-redshift galaxy samples while the blue points represent the binned mean and standard deviation of low-metallicity SDSS galaxies from \protect\cite{Nagao2006}.}
\label{nebmet}
\end{figure*}

\subsubsection{Nebular Line Diagnostics as Metallicity Indicators} 
As discussed earlier, certain nebular line diagnostics may be good indicators of galaxy metallicity \citep{Maiolino2018}.  \cite{Maiolino2008} presents numerous relations between metallicity and various diagnostics for low-redshift SDSS galaxies and it is important to understand whether these relations hold at high-redshift.  In Figure~\ref{nebmet} we plot metallicity (in units of $12+\log_{10}({\rm O/H})$ with ${\rm12+O/H_{\odot}=8.69}$) versus R23, [OIII]~5007$\angstrom/{\rm H\beta}$, and [NII]~6583$\angstrom$/H$\alpha$.  While there is considerable scatter for our three galaxies, it seems that our predicted R23 and [OIII]~5007$\angstrom/{\rm H\beta}$ fall reasonably close to the relations from \cite{Maiolino2008} and the low metallicity SDSS galaxies from \cite{Nagao2006}.  Note that our galaxies fall fairly close to the SDSS tail on the R23-O32 and sSFR-excitation diagrams so there is no reason, {\it a priori}, to expect that these will hugely differ.  However, we find that our galaxies are significantly offset on the [NII]~6583$\angstrom$/H$\alpha$ versus metallicity relation.  We argue that this may be due to modelling inadequacies in the simulations.  Currently, we track only global metallicity rather than abundances of individual elements and thus we have calibrated the x-axis of Figure~\ref{nebmet} to match the oxygen abundance with respect to Solar.  One possible explanation is that we are over-predicting the nitrogen abundance in our galaxies due to our simplistic model for metallicity.  The relationship between N/O and O/H is indeed not constant \citep[e.g.][]{Pilyugin2012}.  Decreasing the nitrogen abundance in our galaxies would also bring our galaxies into better agreement with the $z\sim6$ predictions from \cite{Faisst2016} who used the \cite{Maiolino2008} relations to calibrate metallicity.

One should keep in mind that the enrichment processes in the high-redshift Universe, in particular at $z\sim10$ are different from low redshift.  We expect that enrichment from more exotic processes such at Pop.~III stars and pair-instability SN may impact the elemental abundances in the galaxy.  For the more massive galaxies, Type~II SN will likely be the dominant mechanism of chemical enrichment as the Universe was not yet old enough to have many Type~Ia SN.  These exotic processes can have numerous effects on our diagnostic diagrams such as the BPT diagram.  The harder radiation fields from Pop.~III stars will tend to move the galaxies towards the top right of the BPT diagram (see e.g. \citealt{Kewley2013}).  Furthermore, the yields for oxygen and nitrogen depend on a number of factors such as stellar mass, metallicity, rotation, SN energy, and type of supernova.  This will change the amount of [NII]~6583$\AA$ and [OIII]~5007$\AA$ emission and hence the location of a galaxy on the BPT diagram. The exact yields are still rather uncertain and depend on modelling (see \citealt{Nomoto2013} for more details).  Future simulations will have to capture these effects to better predict how the line diagnostics reflect the metallicity of the high-redshift galaxies.

\section{Discussion and Conclusions}
\label{dc}

We have presented a new suite of cosmological radiation-hydrodynamics simulations, called the Aspen suite, where we used a zoom-in technique to focus our resolution around the environment of a massive LBG in the epoch of reionization.  These simulations have been designed so that nebular and infrared line emission can be computed from the simulations at high accuracy to compare with current and future observations.

The ISM of these galaxies is inhomogeneous in terms of temperature, density, metallicity, ionisation state, ionisation parameter, and the hardness of the radiation field. We have developed a new technique for estimating nebular and infrared line emission from cosmological simulations by training a random forest algorithm (an ensemble machine learning method) on the results of nearly 1~million {\small CLOUDY} models.  We have demonstrated that for nearly all of the bright lines, our method can reproduce the total luminosity to better than 10\% accuracy, which is well within the other systematic uncertainties in calculating this emission.  Furthermore, the new method completes the calculation in a small fraction of the computational time that it would take to run {\small CLOUDY} models for all simulation cells.  

While the Aspen simulations are representative of the state-of-the-art in comparing simulations with high-redshift observations, numerous caveats should be kept in mind when interpreting our results.  In order to achieve the high mass and spatial resolution needed to begin to resolve the properties of the ISM, we have restricted our simulations to a small, and potentially biased, set of galaxies.  Most of our analysis focuses on three massive systems, AD1, AD2, and AD3, and it is not yet clear whether these are representative of the high-redshift galaxy population.  More simulations will be needed to confirm our current results.  Our simulations also employ numerous ``subgrid'' recipes for physics such as SN and metal enrichment.  There is no guarantee that these recipes are truly capturing all of the relevant physics that governs line emission.  Since we use a single scalar to measure local gas-phase metallicity, we do not follow all of the different types of processes that result in abundance ratios that are distinctly non-Solar.  Hence future simulations will need to better capture these effects to improve on our current work.  Furthermore, in many cases, the Stromgren radii of star particles are still unresolved which may impact our ability to model certain phases of the ISM where particular lines originate.  Nevertheless, while there is still much room for improving this current generation of simulations, in the regions of the simulations where the physics is well resolved, we find numerous successes in our ability to explain observations.  In this work, we have systematically compared our simulations with a plethora of high-redshift observations and our main conclusions can be summarised as follows:
\begin{itemize}
\item Infrared [CII]~157.6$\mu$m emission predominantly originates from the cold molecular disk of high-redshift galaxies while [OIII]~88.33$\mu$m emission is concentrated around the young star forming regions.  Surface brightness profiles for both lines tend to be centrally concentrated and extended emission is indicative of a merger.  The time variability of the lines is correlated with how sensitive the morphology of the galaxy is to stellar feedback.

\item Many high-redshift galaxies are expected to exhibit well defined disks and ordered rotation.  This kinematic property is best observable as a velocity gradient in the lines that trace the cold gas disk (such as [CII]~157.6$\mu$m), consistent with the $z\sim6.5$ galaxies observed by \cite{Smit2018}.

\item Kpc-scale spatial offsets between [CII]~157.6$\mu$m and [OIII]~88.33$\mu$m emission, consistent with the observations of \cite{Carniani2017} occur when the galaxy is clumpy and radiation and SN feedback are strong enough to destroy large pockets of neutral gas.  These offsets are not ubiquitous in our simulations.

\item Spectral offsets between [CII]~157.6$\mu$m and [OIII]~88.33$\mu$m emission are naturally produced by our simulation because [OIII]~88.33$\mu$m is very sensitive to the kinematics of individual star forming regions.  [OIII]~88.33$\mu$m spectra often exhibit multiple peaks and tend to have narrower line profiles compared to [CII]~157.6$\mu$m.  The largest spectral offsets of $\gg100$km~s$^{-1}$ are generated in mergers.  

\item Our simulations suggest that massive high-redshift galaxies will fall on the local [CII]-SFR relation, in contrast to previous observations \citep[e.g.][]{Ouchi2013} and simulations \citep[e.g.][]{P17a,P17b}.  We provide evidence that observed deficits may be due to observational selection effects and other biases.

\item Our simulated galaxies show an anti-correlation between [OIII]~88.33$\mu$m/[CII]~157.6$\mu$m and SFR which is consistent with the relation observed between [OIII]~88.33$\mu$m/[CII]~157.6$\mu$m and bolometric luminosity in \cite{Hashimoto2018b,Hashimoto2018c}.  This may be due to the fact that the neutral gas in lower mass galaxies is more susceptible to being disrupted by SN feedback than gas in higher-mass galaxies.

\item The local calibrations \citep{Kennicutt1998a} to estimate SFR from H$\alpha$ and [OII]~3727$\angstrom$ doublet emission provide good fits to our simulated high-redshift galaxies across a broad range in galaxy mass.  We find that H$\beta$ can also be used and provide calibrated relations for all three lines.

\item Our high-redshift galaxies have strong-line diagnostics that are remarkably similar to observed galaxies at $z\sim2-3$ from the KBSS sample \citep{Strom2017}.  We reproduce the offsets on both the BPT and mass-excitation diagrams compared with local galaxies.  Furthermore, our simulated galaxies are also consistent with KBSS with regard to the sSFR-excitation, R23-O32, and O32-Ne3O2 relations providing evidence that the behaviour of the observed high-redshift galaxies on these diagnostic diagrams is driven by high sSFR and excitation and higher ionisation parameters, possibly driven by radiation from low metallicity metal poor binary stars.

\item Although interpreting strong-line diagnostics with a single metallicity and ionisation parameter gives a good intuition of the underlying physics of the ISM, we show that real galaxies have gas that exhibits a huge range in these parameters (varying by more than 10 orders of magnitude in some cases).  Because the ionisation parameter, metallicity, and spectral hardness change drastically between different locations in the same galaxy, we advocate for additional summary statistics that address this diversity of conditions.

\item The local relations from \cite{Maiolino2008} involving R23 and [OIII]~5007$\angstrom$/H$\beta$ can be used as metallicity indicators of high-redshift galaxies.  However, we caution that there is expected to be considerable scatter and our treatment of metallicity in the simulations remains very simplistic.

\end{itemize}
  
Emission lines are an extremely powerful probe of the properties of high-redshift galaxies and here we have expanded on the numerous ways that such observations can be used to elucidate the physics of the ISM in the early Universe.  Being able to properly model the inhomogeneous radiation field, metal enrichment, and ISM are key for interpreting the properties of current and next generation telescopes and here we have provided a step in this direction.  We have shown that by combining the information that can be obtained from IR and nebular lines we can begin to understand the processes governing the earliest generations of star formation in our Universe.  

\section*{Acknowledgements}
This work made considerable use of the open source analysis software {\small PYNBODY} \citep{Pontzen2013}.  We thank the anonymous referee for their detailed revision of the manuscript.  We thank Masami Ouchi, Dan Stark, Akio Inoue, Naoki Yoshida, Kana Moriwaki, Roberto Maiolino, Stefano Carniani, Rebecca Bowler, and Clotilde Laigle for discussions regarding the content of this manuscript.  We thank Allison Strom for providing observational data for KBSS galaxies.  H.K. thanks the Beecroft fellowship, the Nicholas Kurti Junior Fellowship, and Brasenose College. TK was supported by the National Research Foundation of Korea (No. 2017R1A5A1070354 and No. 2018036146). Support by ERC Advanced Grant 320596 ``The Emergence of Structure during the Epoch of reionization" is gratefully acknowledged by MH, HK and TK.  JB and JR acknowledge support from the ORAGE project from the Agence Nationale de la Recherche under grant
ANR-14-CE33-0016-03.  NL and RSE acknowledge funding from the European Research Council (ERC) under the European Union's Horizon 2020 research and innovation programme (grant agreement No 669253).

This work was performed using the DiRAC Data Intensive service at Leicester, operated by the University of Leicester IT Services, which forms part of the STFC DiRAC HPC Facility (www.dirac.ac.uk). The equipment was funded by BEIS capital funding via STFC capital grants ST/K000373/1 and ST/R002363/1 and STFC DiRAC Operations grant ST/R001014/1. DiRAC is part of the National e-Infrastructure.



\bibliographystyle{mnras}
\bibliography{jd1} 

\bsp	
\label{lastpage}
\end{document}